\documentclass[11pt,a4paper]{article}

\usepackage{jheppub}
\pdfoutput=1

\usepackage[utf8]{inputenc}
\usepackage{scalerel}

\usepackage{blindtext}
\usepackage{graphicx}
\usepackage{amsmath,amssymb,multirow,float,bbold,hyperref}
\usepackage{slashed}
\usepackage{color}
\usepackage{mathrsfs}
\usepackage{epsfig}
\usepackage{lipsum}
\usepackage{mathrsfs}
\usepackage{enumerate}
\usepackage[numbers]{natbib}   
\usepackage{dcolumn}
\usepackage{bm}
\usepackage{listings}

\def\beq{\begin{equation}}
\def\eeq{\end{equation}}
\def\bea{\begin{eqnarray}}
\def\eea{\end{eqnarray}}
\def\nn{\nonumber}

\definecolor{gray}{rgb}{0.90,1,1}
\definecolor{LightCyan}{rgb}{0.88,1,1}

\def\bsll{b \to s \ell^+ \ell^-}

\def \cB{{\cal B}}

\def \al{\alpha}

\def \({\left(}
\def \){\right)}
\def \[{\left[}
\def \]{\right]}
\def \l|{\left|}
\def \r|{\right|}

\def \({\left(}
\def \){\right)}
\def \[{\left[}
\def \]{\right]}
\def \l|{\left|}
\def \r|{\right|}
\newcommand{\eps}{\varepsilon}

\def \al{\alpha}

\def\roughly#1{\mathrel{\raise.3ex\hbox
{$#1$\kern-.75em\lower1ex\hbox{$\sim$}}}}
\def\lsim{\roughly<}
\def\gsim{\roughly>}

\title{Dark photon and dark $Z$ mediated $B$ meson decays}

\author[a]{Alakabha Datta,}
\author[b]{A. Hammad,}
\author[c]{Danny Marfatia,}
\author[a]{Lopamudra Mukherjee,}
\author[d]{and Ahmed Rashed}

\affiliation[a]{Department of Physics and Astronomy,
108 Lewis Hall, University of Mississippi, Oxford, MS 38677-1848, USA.}
\affiliation[b]{Institute of Convergence Fundamental Studies, Seoul National University of Science and Technology, 232 Gongneung-ro, Nowon-gu, Seoul, 01811, Korea.}
\affiliation[c]{Department of Physics and Astronomy, University of Hawaii at Manoa,
2505 Correa Rd., Honolulu, HI 96822, USA.}
\affiliation[d]{Department  of Physics, 
Shippensburg University of Pennsylvania, Franklin Science Center, 1871 Old Main Drive, Pennsylvania, 17257, USA.}

\emailAdd{datta@phy.olemiss.edu}
\emailAdd{ahmed.hammad@uniban.ch}
\emailAdd{dmarf8@hawaii.edu}
\emailAdd{lmukherj@olemiss.edu}
\emailAdd{amrashed@ship.edu}

\abstract{
 We study flavor changing neutral current decays of $B$ and $K$ mesons in the dark $U(1)_D$ model, with the dark photon/dark $Z$ mass between 
 10~MeV and 2~GeV. Although the model provides an improved fit (compared to the standard model) to the differential decay distributions of $B \to K^{(*)} \ell^+ \ell^-$, with $\ell= \mu, e$, 
 and $B_s \to \phi \mu^+ \mu^-$,  the allowed parameter space is ruled out by measurements of atomic parity violation, $K^+ \to \mu^+ + invisible$ decay, and $B_s - \overline{B}_s$ mixing, among others. To evade constraints from low energy data, we extend the model to allow for (1) additional invisible $Z_D$ decay, (2) a direct vector coupling of $Z_D$ to muons, and (3) a direct coupling of $Z_D$ to both muons and electrons, with the electron coupling fine-tuned to cancel the $Z_D$ coupling to electrons via mixing. We find that only the latter case survives all constraints.
 } 
 
\begin{document} 
\maketitle

\section{Introduction}
\label{sec:introduction}
 Flavor changing neutral currents (FCNCs) are sensitive probes of new physics (NP) and can strongly constrain many extensions of the standard model (SM). In this work we focus on how the wealth of recent FCNC measurements in the $B$ system constrain models with a light gauge boson. We study the dark $U(1)_D$ model which gives rise to a dark photon or a dark $Z$ in specific circumstances. If only kinetic mixing between  the $U(1)_D$ gauge boson and the electromagnetic field strength tensor occurs, then it is a dark photon with purely vector couplings to all SM fermions except neutrinos~\cite{Holdom:1985ag}.
If the dark vector boson also has mass mixing with SM gauge fields, then it is a dark $Z$, often called a $Z'$~\cite{Gopalakrishna:2008dv, Davoudiasl:2012ag}. We collectively denote a dark photon and a dark $Z$ by $Z_D$. In either case, FCNC processes are generated through loops, and as in the SM, the loops are enhanced for up-type quarks because the large top quark mass suppresses the GIM mechanism. In the case of $D$ decays, the quarks in the loop are down type, which suppresses FCNC decays. Hence, we concentrate on FCNC $B$ and $K$ decays. 

A characteristic feature of models with light mediators is that the new physics cannot be integrated out, resulting in $q^2$ dependent Wilson coefficients (WCs).
The role of light mediators in FCNC $\bsll$ transitions has been discussed extensively~\cite{Datta:2017pfz, Sala:2017ihs, Bishara:2017pje, Ghosh:2017ber, Datta:2017ezo, Altmannshofer:2017bsz, Datta:2018xty, Datta:2019zca,Darme:2020hpo,Borah:2020swo, Darme:2021qzw,Crivellin:2022obd}. In this paper we study a light vector mediator $Z_D$ with mass $0.01 \lsim M_{Z_D}/\text{GeV} \lsim 2$ and allow for onshell as well as off shell effects in $Z_D$ decay. We calculate rates for FCNC processes for both the dark photon and dark $Z$ models.\footnote{Previously in Ref.~\cite{Xu:2015wja}, flavor changing $b \to s Z_D$ processes were studied without including the contribution from the $q^2$ independent dark $Z$ monopole operator. Consequently, $\mathcal{O}(1)$ mixing parameters were incorrectly obtained from $\bsll$ data.} Further, we include hadronic decays of the vector boson in estimating its width. We study extensions of the model with direct interactions of $Z_D$ with muons, and with muons and electrons, apart from mixing induced couplings. We also allow for an additional invisible decay of $Z_D$ which could arise from $Z_D$ couplings to dark sector particles. 

The paper is organized as follows. In Section~2 we discuss the general formalism of FCNC $B$ and $K$ decays in the dark photon and the dark $Z$ models. The Wilson coefficients pertaining to $b \to s \ell^+ \ell^-$ and the treatment of the new physics branching fractions in the narrow width approximation, are provided. In this section, we also calculate the width of the $Z_D$ boson including decays to leptonic, hadronic and invisible states.
In Section~3 we describe our three model cases and place experimental constraints on them in Section~4. In Section~5, we present the results of our fits to $B$ and $K$ decay data. After a short discussion of the muon anomalous magnetic moment in Section~6, we summarize in Section~7.

\section{Formalism}

We assume that $Z_D$ is associated with a broken $U(1)_D$ gauge symmetry of a dark sector and couples with the SM gauge symmetry via kinetic mixing between $U(1)_Y$ and $U(1)_D$ \cite{Holdom:1985ag}.
Following Ref.~\cite{Davoudiasl:2012ag} we write the gauge Lagrangian as,
\bea
{\cal L}_\text{gauge} &= &-\frac{1}{4} B_{\mu\nu} B^{\mu\nu} + \frac{1}{2} \frac{\eps}{\cos\theta_W} B_{\mu\nu} Z_D^{\mu\nu} - \frac{1}{4} Z_{D \mu\nu} Z_D^{\mu\nu}\,, \nonumber\\
B_{\mu\nu} &= & \partial_\mu B_\nu - \partial_\nu B_\mu\,, \qquad Z_{D \mu\nu} = \partial_\mu {Z_D}_\nu - \partial_\nu {Z_D}_\mu\,, \,
\eea
with $\eps$ is a dimensionless parameter and $\theta_W$ is the weak mixing angle.

After diagonalizing the gauge sector~\cite{Chun:2010ve, Davoudiasl:2012ag}, an induced coupling of $Z_D$ to the SM electromagnetic current is generated. To leading order in $\eps$,

\begin{equation}
\mathcal{L}_D^{\text{em}} \supset e \varepsilon Z_D^\mu J_\mu^{\text{em}} -i e \varepsilon \left[\left[Z_D W^+ W^- \right]\right]\,, 
\label{eq:lag1}
\end{equation}
where
\begin{eqnarray}
\left[\left[Z_D W^+ W^- \right]\right] &=&  \varepsilon^\mu_{Z_D}(k_1) \varepsilon^\nu_{W^+}(k_2) \varepsilon^\lambda_{W^-}(k_3) \nn\\
&& \times\left[ (k_1 - k_2)_\lambda g_{\mu\nu} + (k_2 - k_3)_\mu g_{\nu\lambda} + (k_3 - k_1)_\nu g_{\lambda\mu} \right]\,.
\end{eqnarray}
This is the usual {\it dark photon} model. If the $U(1)_D$ symmetry is broken by a scalar that is charged under the SM, then $Z_D$ can mix with the SM $Z$ boson via mass terms~\cite{Gopalakrishna:2008dv, Davoudiasl:2012ag}. The physical eigenstates can be written in terms of the weak eigenstates as
\bea
Z   &=  & Z^0 \cos\xi - Z_D^0 \sin\xi\,,  \nonumber\\
Z_D &=  & Z^0 \sin\xi + Z_D^0 \cos\xi\,, \
\label{eq:mass-mixing}
\eea
where $\xi$ parameterizes the mass mixing between the gauge bosons. This induces a coupling of $Z_D$ with the SM fields given by,
\begin{equation}
\mathcal{L}_D^{Z} \supset \frac{g}{\cos \theta_W} 
\varepsilon_Z Z_D^\mu J_\mu^{\text{Z}}-i g \cos \theta_W\varepsilon_Z \left[\left[Z_D W^+ W^- \right]\right]\,,
\label{eq:lag2}
\end{equation}
where $\eps_Z \equiv {1\over 2}\tan 2\xi$.
Such an interaction defines the {\it dark $Z$} model. Note that if $U(1)_D$ is broken by SM singlet scalars, then $\eps_Z=0$ and we recover the special case of a dark photon.
We do not consider a specific Higgs sector for the spontaneous symmetry breaking of $U(1)_D$, and instead study mass mixing of the general form in Eq.~(\ref{eq:mass-mixing}). Since the gauge kinetic mixing parameter $\varepsilon$ affects the mass-mixing angle $\xi$ at subleading order, we neglect its effect. The free parameters of the model are the mixing parameters $\varepsilon$ and $\varepsilon_Z$, and the mass of the dark boson $M_{Z_D}$. 
Updated allowed values of these parameters can be found in Refs.~\cite{Bertuzzo:2018ftf,Bertuzzo:2018itn,Link:2019pbm}. 


\subsection{Partonic amplitude of FCNC processes}

\begin{figure}[t]
\begin{center}
\includegraphics[width=13cm]{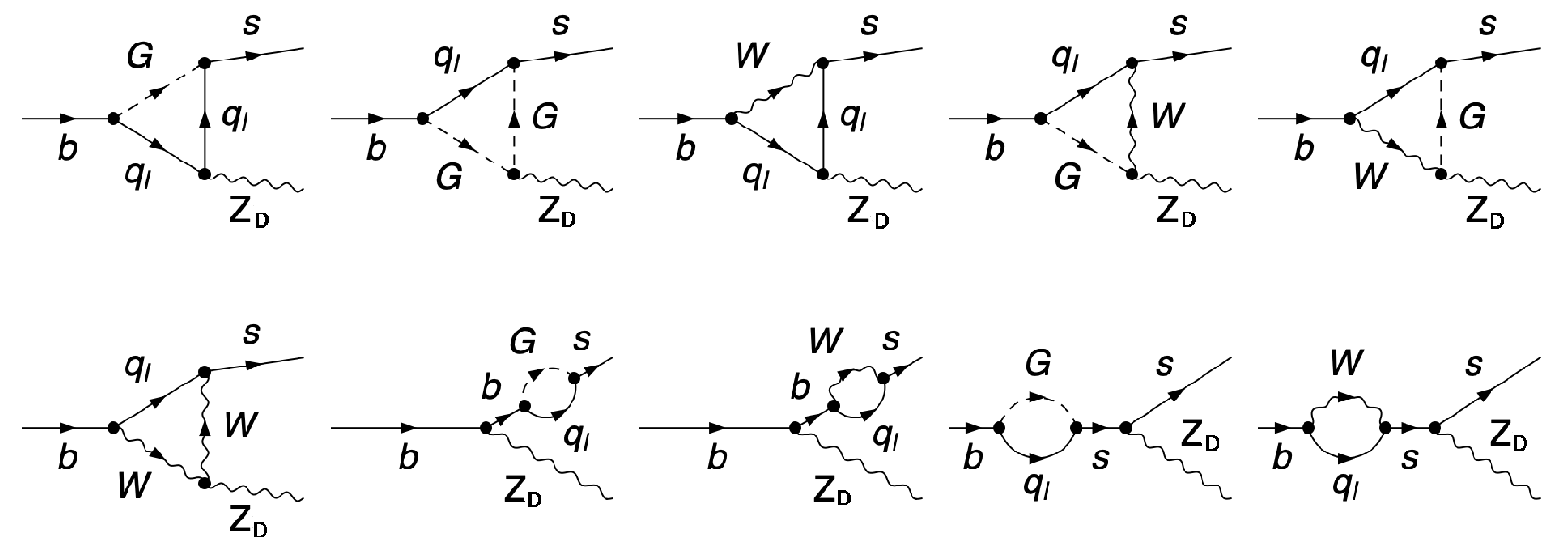}
\end{center}
\caption{Feynman diagrams for the FCNC process $b\to sZ_D$ at the parton level. $G$ denotes Goldstone bosons.}
\label{fig:loop-diags}
\end{figure}

FCNC $B$ and $K$ decays  proceed via the underlying quark level transitions $ d_i \to d_j Z_D$, where $d_{i,j}$ are down type quarks. The quark level transitions are $b \rightarrow s$, $b \rightarrow d$ and $s \rightarrow d$. Since $Z_D$ is flavor conserving, the decay occurs at loop level. 
The process $d_i\rightarrow d_j Z_D$ can be written in terms of the SM $d_i\rightarrow d_j \gamma$ and $d_i\rightarrow d_j Z$ processes by modifying the couplings to include the kinetic and mass mixing parameters $(\varepsilon,\; \varepsilon_Z)$. The processes can be described by generic effective local operators and their Wilson coefficients. The monopole operators, which conserve chirality, are 
\begin{eqnarray}
\label{e0ande2}
H_{\rm eff} \supset 
\left(\bar F_a^\prime \gamma^\mu P_{L,R}F_b\right)\left[\left(E^{0,c}_{a,b}\right)_{L,R}
g^{\mu\nu}
+ 
\left(g^{\mu\nu}q^2-q^\mu q^\nu\right)\left(E^{2,c}_{a,b}\right)_{L,R}
\right]V^c_\nu\,,
\label{eq:monopole}
\end{eqnarray}
and the dipole operators, which flip chirality,  are
\begin{eqnarray}
\label{eq:dipole}
H_{\rm eff} \supset
\left(\bar F_a^\prime \sigma^{\mu\nu} P_{L,R}F_bq_\mu V^c_\nu\right)\left(M^{1,c}_{a,b}\right)_{L,R}\,,
\end{eqnarray}
where  $P_{L,R}=(1/2)(1\mp\gamma^5)$ are  the projection operators, the metric tensor is defined as $g^{\mu\nu}=\mathrm{diag}(1,-1,-1,-1)$, $\sigma^{\mu\nu}\equiv (i/2)\left[\gamma^\mu,\gamma^\nu\right]$, and $q^\mu$ is the outgoing momentum of a neutral vector boson $V$. Fermions of different families are denoted by $F$ and $F^\prime$, and $a,~b,~c$ are color indices. Note that in Ref.~\cite{Xu:2015wja}, the first term in Eq.~\eqref{eq:monopole} which provides the leading contribution for a  dark $Z$, is neglected.

In the above basis of the effective Hamiltonian, the hadronic part of the amplitude involving different Lorentz structures is given by
\begin{eqnarray}
\mathcal{M}_{Z_D} &=&  \langle \mathcal{H}_2 | \bar d_i \gamma_\mu P_{L/R} d_j | \mathcal{\bar H}_1   \rangle  \left[\lbrace\left(E^{0,A}_{c_1,c_2}\right)_{L/R}+\left(E^{0,Z}_{c_1,c_2}\right)_{L/R}\rbrace g^{\mu\nu}\right. \nonumber\\
&+& \left.\lbrace\left(E^{2,A}_{c_1,c_2}\right)_{L/R}+\left(E^{2,Z}_{c_1,c_2}\right)_{L/R}\rbrace \left(g^{\mu\nu}q^2-q^\mu q^\nu\right)\right]V_\nu^{Z_D} \nonumber\\
&+&   \langle \mathcal{H}_2 | \bar d_i iq_\mu \sigma^{\mu\nu} P_{L/R} d_j | \mathcal{\bar H}_1 \rangle  \lbrace\left(M^{1,A}_{c_1,c_2}\right)_{L/R}+\left(M^{1,Z}_{c_1,c_2}\right)_{L/R}\rbrace V_\nu^{Z_D}\,,
\label{DarkPhotonZ}
\end{eqnarray}
with $V_\nu^{Z_D}$ is the polarization vector of $Z_D$ and the hadronic currents are provided in Appendix~\ref{sec:FF}. We use the {\tt Peng4BSM@LO} package~\cite{Bednyakov:2013tca} to calculate the amplitudes of $b\rightarrow s Z_D,\; s \to d Z_D,$ and $b\rightarrow d Z_D$ that arise from penguin diagrams; see Fig.~\ref{fig:loop-diags}. 
The loop functions are calculated to first order in the small masses and momenta of the external fermions. 
The contribution of the monopole terms dominates over the dipole terms. Expressions for the various loop factors of the triangle diagrams are given in Appendix~\ref{sec:loop-factors}. 

\subsection{Semileptonic FCNC amplitudes}
Semileptonic $\bsll$ processes are described by the effective Hamiltonian,
\beq
\mathcal{H}_{\text{eff}}^{bsll} = -\frac{4G_F}{\sqrt{2}}\frac{e^2}{16\pi^2}V_{tb}V_{ts}^* \sum_i (\mathcal{C}_i \mathcal{O}_i + \mathcal{C}_i^\prime \mathcal{O}_i^\prime)\,,
\eeq 
where $\mathcal{C}_i^{(\prime)}$ are the WCs corresponding to the dimension six operators $\mathcal{O}_i^{(\prime)}$. The  operators relevant to our study are $\mathcal{O}_9 = (\bar{s}_L \gamma^\mu b_L)(\bar{\ell} \gamma_\mu \ell)$ and $\mathcal{O}_{10} = (\bar{s}_L \gamma^\mu b_L)(\bar{\ell} \gamma_\mu \gamma_5 \ell)$ and their primed counterparts, which are obtained by flipping the chirality $L \to R$.

To calculate the WC for the $\bsll$ transition, we combine the hadronic loop factors of the previous section, with the appropriate leptonic $Z_D \ell \ell$ vertex factor and the $Z_D$ propagator, to get
\bea
\mathcal{C}_{9,\ell} &=&\left[\left(\left(E^{0,Z}_{c_1,c_2}\right)_{L} + \lbrace\left(E^{2,A}_{c_1,c_2}\right)_{L}+\left(E^{2,Z}_{c_1,c_2}\right)_{L}\rbrace q^2 \right) \right. \nonumber\\
& \times &  \left. \left(\frac{1}{q^2 - M_{Z_D}^2 + i \Gamma_{Z_D} M_{Z_D}}\right) \left(e\eps + \frac{g}{c_W}\eps_Z g_V^\ell \right) \right]\,, \label{c9}\\
\mathcal{C}_{10,\ell} &=& \left[\left( \left(E^{0,Z}_{c_1,c_2}\right)_{L} + \lbrace\left(E^{2,A}_{c_1,c_2}\right)_{L}+\left(E^{2,Z}_{c_1,c_2}\right)_{L}\rbrace q^2 \right) \right. \nonumber\\
& \times &  \left. \left(\frac{1}{q^2 - M_{Z_D}^2 + i \Gamma_{Z_D} M_{Z_D}}\right) \left(\frac{g}{c_W}\eps_Z g_A^\ell \right) \right]\,, \label{eq:WC-NP}\\
\mathcal{C}_{9,\ell}^{\prime} &=& \mathcal{C}_{10,\ell}^{\prime} = 0\,,
\eea
where $g_V^\ell = (-1 + 4 s^2_W)/2$ and $g_A^\ell = -1/2$ are the vector and axial vector coupling constants for the SM $Z\ell \ell$ interaction; $s_W$ and $c_W$ are the sine and cosine of $\theta_W$, respectively.  When obvious, we suppress the 
$\ell$ subscript in the WCs.

If $M_{Z_D}$ is outside the measured $q^2$ bin, an off-shell contribution to the semileptonic decay amplitude can interfere effect with the SM amplitude. However, if $M_{Z_D}$ lies within a particular $q^2$ bin, then the new physics contribution can be treated as arising from the on-shell production of the light boson followed by its subsequent decay to lepton pairs ($\mathcal{H}_1 \to \mathcal{H}_2 Z_D\,, Z_D \to \ell^+ \ell^-$). As the width of the light boson is much smaller than the energy resolution of the detector, the narrow width approximation can be employed to write the branching fraction as~\cite{Altmannshofer:2017bsz}
\beq
\langle\cB\rangle_{b\to s \ell\ell}|_{q_{min}}^{q_{max}}
= \langle\cB\rangle_{b\to s \ell\ell}^{SM}|_{q_{min}}^{q_{max}} + \cB(\mathcal{H}_1 \to \mathcal{H}_2 Z_D)\cB(Z_D \to \ell^+\ell^-)\cdot \mathcal{G}^{(r_\ell)}(q_{min},q_{max})\,,
\eeq
where
\beq 
\mathcal{G}^{(r_\ell)}(q_{min},q_{max})=\frac{1}{\sqrt{2\pi} r_\ell}\int_{q_{min}}^{q_{max}}d|q| e^{-\frac{\left(|q|-M_{Z_D}\right)^2}{2r_\ell^2}}\,,
\eeq 
is a Gaussian smearing function with $r_e = 10$~MeV and $r_\mu = 2$~MeV. It is important to note that the on-shell decay of the light boson always enhances the contribution to the branching fraction in a bin. 

\subsection{Dark boson decay width}

In the mass range of interest, $Z_D$ decays to lepton pairs and to hadronic final states. While the $e^+e^-$ and $\nu\bar\nu$ final states are always kinematically allowed, the $\mu^+ \mu^-$ final state is possible only for $M_{Z_D} > 2 m_\mu$. The decay widths to charged and neutral leptons are given by
\begin{eqnarray}
\Gamma(Z_D \rightarrow \ell^+ \ell^-) &=& \frac{\text{e}^2  }{96 \pi  \text{$c_W^2$} \text{$s_W^2$} \text{$M_{Z_D}$}}\sqrt{ 1-4\frac{\text{$m_\ell^2$}}{\text{$M_{Z_D}^2$}}}\bigg(8 \text{$c_W^2$} \text{$s_W^2$} \eps^2 (2 \text{$m_\ell^2$}+\text{$M_{Z_D}^2$})\nonumber\\
&&-4 \text{$c_W$} \text{$s_W$} (4 \text{$s_W^2$}-1) \eps \text{$\eps_Z$} (2 \text{$m_\ell^2$}+\text{$M_{Z_D}^2$})
\nonumber\\
&&+\text{$\eps_Z^2$} \big(\text{$m_\ell^2$} (16 \text{$s_W^4$}-8 \text{$s_W^2$}-1)+\text{$M_{Z_D}^2$} (8 \text{$s_W^4$}-4 \text{$s_W^2$}+1)\big)\bigg)\,,\\
\Gamma(Z_D \rightarrow \nu \bar{\nu})&=&\frac{\text{$e^2$} \text{$M_{Z_D}$} \text{$\eps_Z^2$}}{96 \pi  \text{$c_W^2$}  \text{$s_W^2$}}\,.
\end{eqnarray}

\begin{figure}[t]
    \centering
    \includegraphics[scale=0.75]{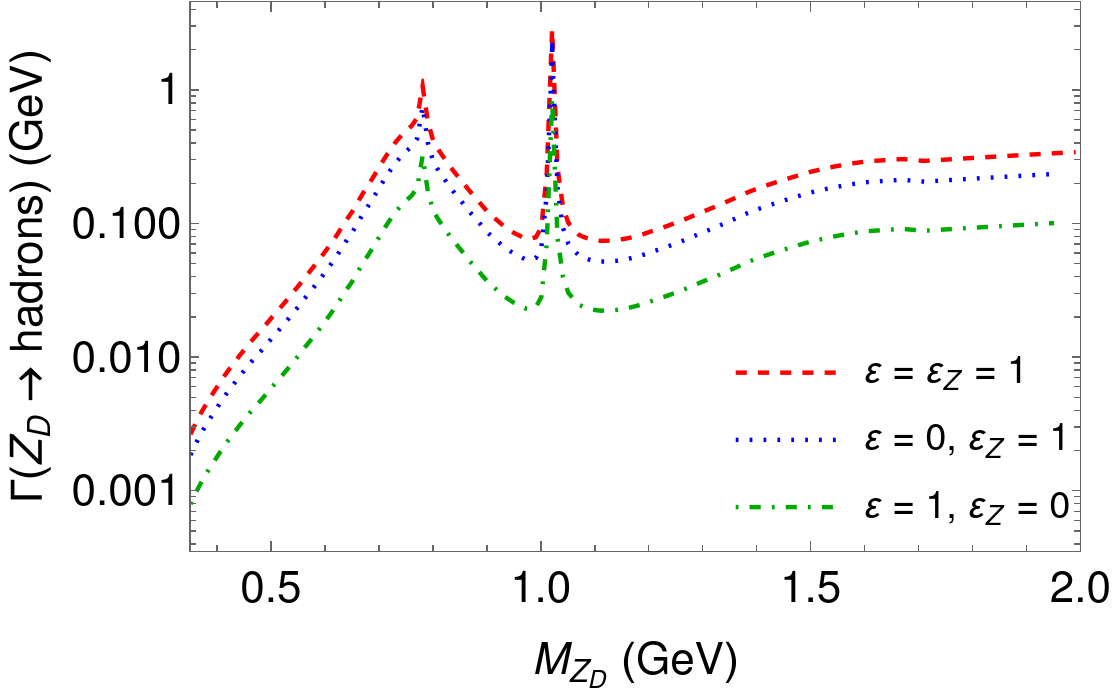}
    \caption{Decay width of the dark boson to light hadronic final states for different mixing parameter values.}
    \label{fig:ZD-hadron}
\end{figure}

Hadronic decays, however, cannot be calculated using perturbative QCD and we rely on the vector meson dominance model (VMD) to describe low energy QCD~\cite{Sakurai:1960ju, Kroll:1967it, Lee:1967iv, Fraas:1969uwg, Bando:1984ej}. Recently, the hadronic decay width of light $U(1)$ vector bosons has been calculated in a data driven approach~\cite{Tulin:2014tya, Ilten:2018crw, Foguel:2022ppx}. The hadronic decay of a dark photon, with coupling proportional to the electric charge,  can be expressed by an appropriate rescaling as
\beq
\Gamma(Z_D \to \mathcal{H}) = \Gamma(Z_D \to \mu^+ \mu^-) \times \mathcal{R}^\mathcal{H}_\mu
\eeq 
where $\mathcal{R}^\mathcal{H}_\mu = \sigma(e^+ e^- \to \mathcal{H})/\sigma(e^+e^- \to \mu^+ \mu^-)$ has been experimentally measured~\cite{ParticleDataGroup:2022pth}. For energies far from the hadron resonances, $e^+e^-$ annihilation slowly transitions towards perturbative quark pair production, in which case $\mathcal{R}^\mathcal{H}_\mu \simeq N_c \sum_{f=u,d,s} (Q_{em}^f)^2 = 2$ for SM, where $N_c = 3$ is the color factor and $Q_{em}^f$ is the electromagnetic charge of fermion $f$. The hadronic decay width of baryophilic dark photons has been computed in Ref.~\cite{Foguel:2022ppx} by summing over various hadronic final states in the VMD model. We adapt this analysis to the vector coupling of the dark $Z$. For the axial vector coupling of $Z_D$ we use quark level decays to estimate the contribution to the hadronic width. In Fig.~\ref{fig:ZD-hadron} we show the $Z_D$ partial width to hadronic final states for different mixing parameter values.

\section{Models}
\label{sec:meth}
We study three different cases of the light $Z_D$ model as specified below.
\begin{itemize}
    \item {\textbf{Case A}}: This is the dark photon and dark $Z$ model described by Eqs.~\eqref{eq:lag1} and~\eqref{eq:lag2}. The model has two mixing parameters ($\eps$ and $\eps_Z$) and the mass $M_{Z_D}$.
    \item {\textbf{Case B}}: A muonphilic $Z_D$ in which Case A is extended with an additional direct interaction of the dark $Z$ with muons:
    \beq 
\mathcal{L}_D^Z \supset g_{D}^\mu \bar{\mu} \gamma_\al \mu Z_{D}^\al\,.
\label{eq:CaseB}
\eeq 
This scenario has an additional free parameter $g_{D}^\mu$.
    \item {\textbf{Case C}}: Case A is extended with additional direct interactions of the dark $Z$ with both electrons and muons: 
    \beq 
\mathcal{L}_D^Z \supset g_{D}^e \bar{e} \gamma_\al e Z_{D}^\al + g_{D}^\mu \bar{\mu} \gamma_\al \mu Z_{D}^\al\,.
\label{eq:CaseC}
\eeq
We assume $g_{D}^e$ is fine-tuned so that it cancels the coupling of $Z_D$ to electrons via mixing. Then, all observables for the electron mode are described by the SM only.
\end{itemize}

\section{Constraints}

\subsection{$B_s$ mixing}

$B$ meson mixing plays an important role in the search for new physics and provides a strong constraint on our models. In the SM, the mixing has its origin in a box diagram with a 
$W$~boson and top quark in the loop. The dominant contribution to the mass difference due to $B_s^0 - \bar{B}_s^0$ mixing is given by~\cite{Artuso:2015swg}
\begin{equation}
 \Delta M_{B_s}^{SM} = \frac{G_F^2 M_{B_s}}{6\pi^2} M_W^2 (V_{tb} V_{ts}^{*})^2 \eta_B f_{B_s}^2 \hat B_{B_s} S_0(\overline{m}_t^2/M_W^2)\,,
 \label{eq:delMs-SM}
\end{equation}
where $S_0(x)$ is the Inami-Lim function~\cite{Inami:1980fz},
\beq 
S_0 (x) = \frac{4x - 11x^2 +x^3}{4(1-x)^2} - \frac{3x \ln x}{2(1-x)^2}\,,
\eeq
and $\overline{m}_t = 163.53$~GeV is the $\overline{MS}$ mass of the top quark~\cite{FermilabLattice:2016ipl}. Here, $\eta_B \approx 0.551$ is a numerical factor that arises from the leading and next-to-leading order QCD corrections to the box diagram~\cite{Buchalla:1995vs}. The long-distance physics in the hadronic matrix element,  parametrized by the decay constant $f_{B_s}$ and the bag parameter $\hat B_{B_s}$, is the major source of theoretical uncertainty. The product $f_{B_s} \sqrt{\hat B_{B_s}} = 256.1(5.7)$~MeV is obtained from lattice calculations with $N_f = 2+1+1$ dynamical fermions~\cite{Aoki:2021kgd}. The dominant NP correction to the mass difference comes from the monopole operator and is 
given by~\cite{Datta:2017ezo}
\begin{equation}
 \Delta M_{B_s}^{NP} = \frac{1}{3}\frac{1}{M_{B_s}^2 - M_{Z_D}^2}  f_{B_s}^2 \hat B_{B_s} M_{B_s} \lbrace(E^{0,Z}_{b,s})_L\rbrace^2 \left(1-\frac{5}{8}\frac{m_b^2}{M_{Z_D}^2}\right)\,.
 \label{eq:delMs-NP}
\end{equation}
Note that $(E^{0,Z}_{b,s})_L$ does not depend on $\eps$ at one-loop order.

In the SM, the mass difference is~\cite{DiLuzio:2019jyq}
\beq
\Delta M_{B_s}^{SM} = (18.4^{+0.7}_{-1.2}) \text{ ps}^{-1}\,,
\label{eq:deltaMs-SM-prediction}
\eeq
which is in agreement with the experimental measurement~\cite{ParticleDataGroup:2022pth}
\beq
\Delta M_{B_s}^{exp} = (17.765 \pm 0.006) \text{ ps}^{-1}\,.
\label{eq:deltaMs-exp-prediction}
\eeq

\begin{figure}[t]
    \centering
    \includegraphics[scale=0.85]{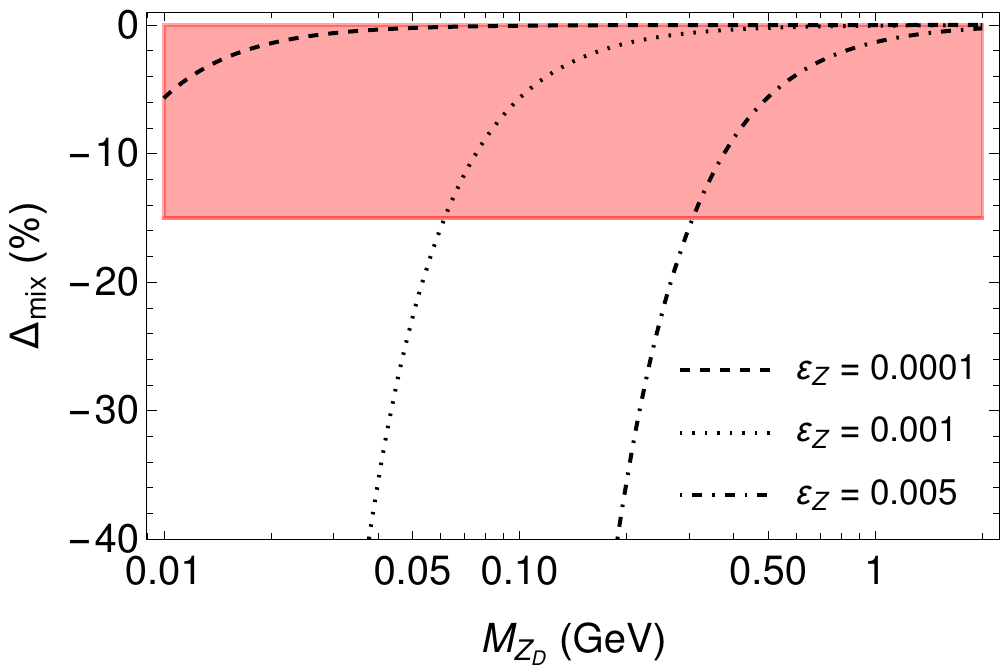}
    \caption{Sensitivity of $B_s^0-\overline{B}_s^0$ mixing to $\eps_Z$ as a function of $M_{Z_D}$. At leading order $\Delta_{mix}$ is independent of $\eps$. The red band is the uncertainty in $\Delta_{\rm mix}$ taken to be the $2\sigma$ lower uncertainty in $\Delta M_{B_s}^{SM}$.}
    \label{fig:Bs-mixing-CaseA}
\end{figure}

The total mass difference including the $Z_D$ contribution can be expressed as
\beq
\Delta M_{B_s} = \Delta M_{B_s}^{SM} + \Delta M_{B_s}^{NP} = \Delta M_{B_s}^{SM} \left(1 + \frac{\Delta M_{B_s}^{NP}}{\Delta M_{B_s}^{SM}}\right) \equiv \Delta M_{B_s}^{SM} (1 + \Delta_{mix})\,,
\eeq
where $\Delta_{mix}$ contains the NP information and is free from uncertainties in the decay constant and bag factor.
Since $\Delta_{mix} \leq 0$, the 6.5\% lower uncertainty in the SM expectation translates into a 6.5\% uncertainty in $\Delta_{mix}$. We plot $\Delta_{mix}$ as a function of the dark $Z$ mass for different values of $\eps_Z$ in Fig.~\ref{fig:Bs-mixing-CaseA}. It is evident that lighter $Z_D$ require smaller values of $\eps_Z$ for $\Delta_{mix}$ to lie within the $2\sigma$ uncertainty of the SM prediction. We find that $\eps_Z \gsim 0.001$ is disallowed for $M_{Z_D} \lsim 60$~MeV.

\subsection{$B_s \to \mu^+ \mu^-$}
The branching fraction of the rare $B_s \to \mu^+\mu^-$ decay, including new physics contributions, is given by
\begin{align}
    \mathcal{B}(B_s \to \mu^+ \mu^-) = \tau_{B_s} f_{B_s}^2 m_{B_s}^3 &\frac{G_F^2 \alpha^2}{64\pi^3} |V_{tb}V_{ts}^*| \sqrt{1-\frac{4m_\mu^2}{m_{B_s}^2}}~\bigg[\frac{m_{B_s}^2}{m_b^2}\left(1-\frac{4m_\mu^2}{m_{B_s}^2}\right)|\mathcal{C}_S -\mathcal{C}_S^\prime|^2 \nonumber \\
    & + \left|\frac{m_{B_s}}{m_b}(\mathcal{C}_P -\mathcal{C}_P^\prime) + 2\frac{m_\mu}{m_{B_s}} (\mathcal{C}_{10}-\mathcal{C}_{10}^\prime)\right|^2\bigg]\,.
\end{align}
In the $Z_D$ model, contributions from the scalar and pseudoscalar WCs, $\mathcal{C}_S^{(\prime)}$ and $\mathcal{C}_P^{(\prime)}$, are absent. The dominant contribution therefore comes from $\mathcal{C}_{10}$ as given in Eq.~\eqref{eq:WC-NP}. For our numerical estimates we use $f_{B_s} = 230.3(1.3)$ MeV~\cite{FlavourLatticeAveragingGroup:2019iem}.
The SM expectation~\cite{Altmannshofer:2021qrr, Guadagnoli:2022izc}  and experimental measurements are~\cite{LHCb:2021awg} 
\bea 
\mathcal{B}(B_s^0 \to \mu^+ \mu^-)^{SM} &=& (3.67 \pm 0.15)\times 10^{-9} \label{eq:Bsmumu-SM}\,,\\
\mathcal{B}(B_s^0 \to \mu^+ \mu^-)^{LHCb~2021} &=& (3.09^{+0.46+0.15}_{-0.43-0.11})\times 10^{-9}\,.
\label{eq:Bsmumu-exp}
\eea

\begin{figure}[t]
    \centering
    \includegraphics[scale=0.85]{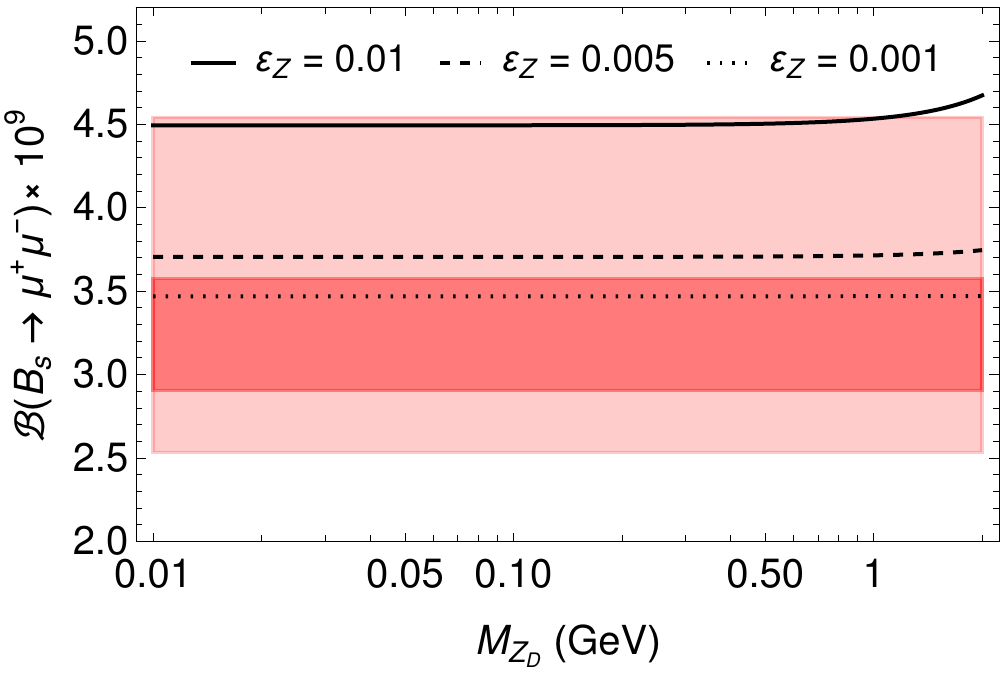}
    \caption{Branching fraction for $B_s \to \mu^+ \mu^-$ for different values of $\varepsilon_Z$. The horizontal red (light red) band denotes the $1\sigma\, (3 \sigma)$ allowed region from experiment~\cite{LHCb:2021awg}. The decay rate does not depend on $\eps$.} 
    \label{fig:BsMuMu}
\end{figure}

The $Z_D$ contribution to this rare decay is shown in Fig.~\ref{fig:BsMuMu}. Since the decay rate depends only on the new axial-vector interaction of the dark $Z$, it is independent of 
$\eps$. It is evident that $\eps_Z$ as large as $0.01$ is allowed by the data at the $3\sigma$ confidence level (CL). 

\subsection{$B \to K^{(*)} \nu \bar{\nu}$}

$B \to K^{(*)} \nu \bar{\nu}$ can be considered as a resonance decay with on-shell production of $Z_D$ from $B$ decay, followed by $Z_D \to \nu \bar{\nu}$. The decay widths for dark $Z$ production can be written as~\cite{Fuyuto:2015gmk}
\beq
    \Gamma(B \to K Z_D) = \frac{|g_V^\nu|^2}{64\pi} \frac{\lambda_K^{3/2}}{m_B^3 M_{Z_D}^2}|f_+(M_{Z_D}^2)|^2\,,
\eeq 
\beq
    \Gamma(B \to K^* Z_D) = \frac{\lambda_{K^*}^{1/2}}{16\pi m_B^3}(|H_0|^2 + |H_+|^2 + |H_-|^2)\,,
\eeq 
where $\lambda_{K^{(*)}}$ stands for the Kallen function,
\beq
    \lambda_{K^{(*)}}(q^2) = m_B^4 + m_{K^{(*)}}^4 + q^4 - 2 (m_B^2 m_{K^{(*)}}^2 + m_{K^{(*)}}^2 q^2 + m_B^2 q^2)\,,
\eeq 
and the helicity amplitudes are given by
\begin{eqnarray}
    H_0 &=& g_A^\nu \left(-\frac{1}{2}(m_B+m_{K^*}) A_1(M_{Z_D}^2)x_{K^* Z_D} + \frac{m_{K^*} M_{Z_D}}{m_B + m_{K^*}}A_2(M_{Z_D}^2) (x_{K^* Z_D}^2 -1)\right),\nonumber\\
    H_\pm &=& \frac{g_A^\nu}{2}(m_B+m_{K^*})A_1(M_{Z_D}^2) + g_V^\nu \frac{m_{K^*} M_{Z_D}}{m_B + m_{K^*}} V(M_{Z_D}^2)\sqrt{x_{K^* Z_D}^2 -1}\,,
\end{eqnarray}
with $x_{K^*{Z_D}} = (m_B^2 - m_{K^*}^2 - M_{Z_D}^2)/(2m_K^* M_{Z_D})$. The effective interaction strengths $g_V^\nu, g_A^\nu$ are related to the loop functions by $g_{V(A)}^\nu = (E^{0,Z}_L \pm E^{0,Z}_R)$. The form factors $V(q^2), A_1(q^2)$ and $A_2(q^2)$ can be found in Ref.~\cite{Ball:2004rg}.

\begin{figure}[t]
    \centering
    \includegraphics[scale=0.65]{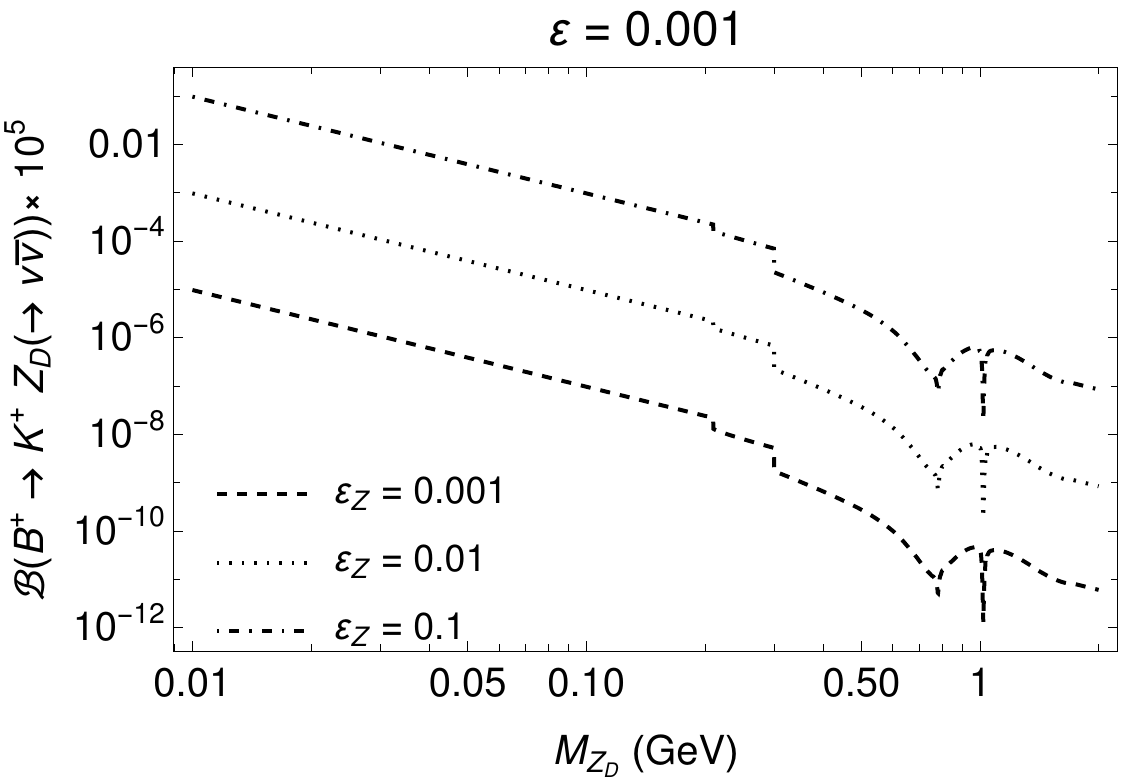}
    \includegraphics[scale=0.65]{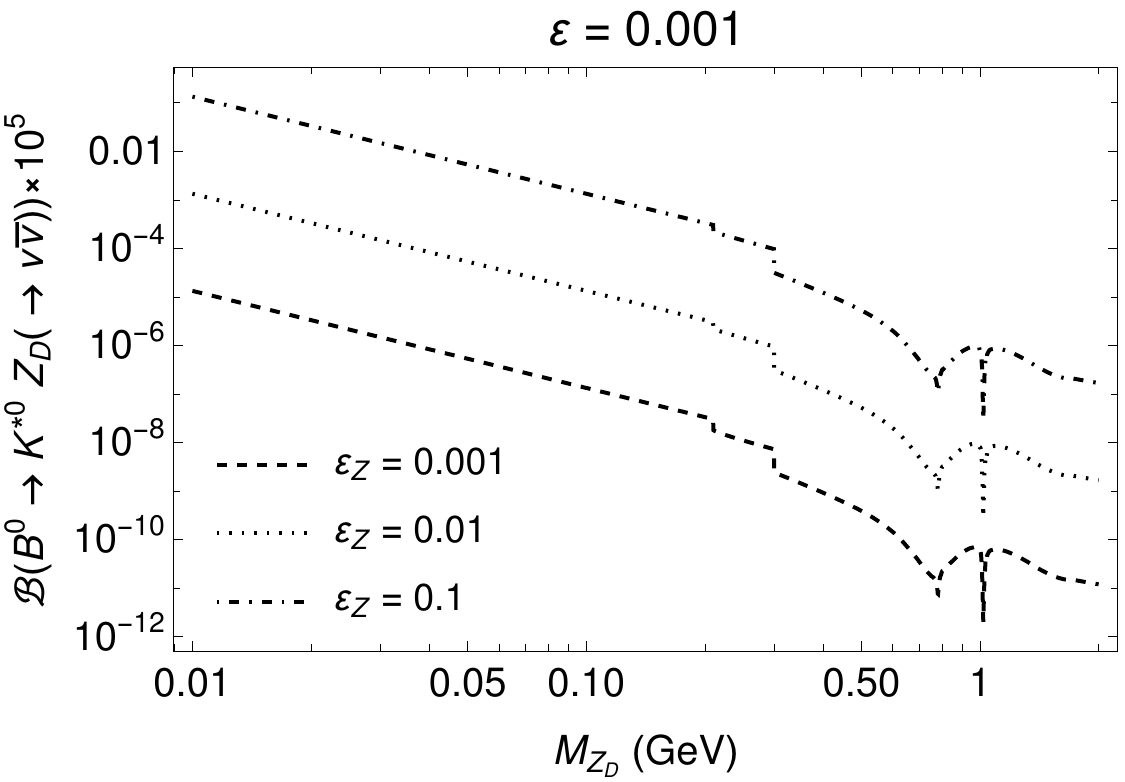}
    \caption{Branching fraction of $B \to K^{(*)} \nu \bar\nu$ as a function of the $Z_D$ mass for three values of $\eps_Z$ and $\eps=0.001$.}
    \label{fig:BtoKnunu}
\end{figure}

The SM predictions for the above flavor changing decays are~\cite{Felkl:2021uxi}
\bea 
\mathcal{B}(B^+ \to K^+ \nu \bar{\nu})_{SM} &=& (4.4 \pm 0.7) \times 10^{-6}\,, \\
\mathcal{B}(B^0 \to K^{*0} \nu \bar{\nu})_{SM} &=& (11.6 \pm 1.1) \times 10^{-6}\,.
\label{eq:BKKstnunu-SM}
\eea
A recent search for $B^+ \to K^+ \nu \bar{\nu}$ by the Belle~II experiment using a new inclusive tagging method led to the $90\%$ CL upper bound~\cite{Dattola:2021cmw},
\beq 
\mathcal{B}(B^+ \to K^+ \nu \bar{\nu}) < 4.1 \times 10^{-5}\,,
\label{eq:BKnunu-BelleII}
\eeq 
which, when combined with previous measurements by Belle and Babar puts the weighted average at
\beq 
\mathcal{B}(B^+ \to K^+ \nu \bar{\nu})_{WA} = (1.1 \pm 0.4) \times 10^{-5}\,.
\label{eq:BKnunu-WA}
\eeq
Although the weighted average shows an enhancement over the SM expectation, caution should be exercised before treating it as a hint of new physics~\cite{Felkl:2021uxi}.
The most recent 90\%~CL upper bound on the branching fraction of $B \to K^* \nu \nu$ is~\cite{Belle:2017oht}
\beq 
\mathcal{B}(B^0 \to K^{*0} \nu \bar{\nu}) < 1.8 \times 10^{-5}\,.
\label{eq:BKstnunu-Belle}
\eeq
In Fig.~\ref{fig:BtoKnunu}, we plot the branching fractions for some benchmark values of $\eps_Z$. We find that they are more than an order of magnitude smaller than the respective upper bounds even for $\eps_Z$ as large as 0.1. 

\subsection{Kaon decay and mixing}
The flavor changing $K \to \pi \nu \bar \nu$ decay is  mediated by the $s \to d \nu \bar{\nu}$ transition at the quark level. The two golden decay modes are $K^+ \to \pi^+ \nu \bar{\nu}$ and $K_L \to \pi^0 \nu \bar{\nu}$. The most recent measurement of the branching fraction of the charged kaon decay is that of the NA62 experiment~\cite{NA62:2021zjw},
\beq
    \mathcal{B}(K^+ \to \pi^+ \nu \bar{\nu}) = (10.6^{+4.0}_{-3.4}|_{stat}\pm 0.9_{sys})\times 10^{-11}\,,
\eeq
while the 90\% CL upper bound on $K_L$ decay from the KOTO experiment~\cite{KOTO:2020prk} is
\beq
 \mathcal{B}(K_L \to \pi^0 \nu \bar{\nu}) < 4.9 \times 10^{-9}\,.
\eeq 
The SM expectations for these decays are~\cite{Buras:2015qea, PDG2022}
\bea
\mathcal{B}(K^+ \to \pi^+ \nu \bar{\nu})_{SM} &=& (8.4 \pm 1.0) \times 10^{-11}\,, \\
\mathcal{B}(K_L \to \pi^0 \nu \bar{\nu})_{SM} &=& (3.4 \pm 0.6) \times 10^{-11}\,.
\eea 
The two decays are theoretically related in a model-independent manner by the Grossman-Nir bound~\cite{Grossman:1997sk} via $\mathcal{B}(K_L \to \pi^0 \nu \bar{\nu}) \lsim 4.3~\mathcal{B}(K^+ \to \pi^+ \nu \bar{\nu})$.

As in the case of $B \to K \nu \nu$, the dark $Z$ contribution to kaon decay can be considered to be resonant. The hadronic part of the amplitude is similar to that in Fig.~\ref{fig:loop-diags} with $s$ and $d$ quarks in the external legs. The $s \to d Z_D$ transition is much suppressed compared to $b \to s Z_D$ and so the branching fractions of semileptonic kaon decays to invisible final states are tiny even for $\eps_Z \sim 0.01$. 

The neutral kaons $K^0$ and $\overline{K}^0$ undergo oscillations with a corresponding mass splitting just like the $B_s^0$ mesons. The leading SM contribution to the mass difference is given by
\begin{equation}
 \Delta M_{K}^{SM} = \frac{G_F^2 M_{K}}{6\pi^2} M_W^2 (V_{cd} V_{cs}^{*})^2 \eta_1 f_{K}^2  \hat B_{K} S_0( m_c^2/M_W^2) \,,
 \label{eq:delMK-SM}
\end{equation}
where $\eta_1 = 1.38$ is the QCD correction factor, $f_K = 155.7(0.3)$ MeV \cite{Aoki:2021kgd} is the kaon decay constant, and $\hat{B}_K = 0.717(18)(16)$~\cite{Aoki:2021kgd} is the bag factor. The NP contribution is determined by the monopole operator and is given by
\beq 
\Delta M_{K}^{NP} = \frac{1}{3}\frac{1}{M_{K}^2 - M_{Z_D}^2}  f_{K}^2 \hat B_{K} M_{K} (E^{0,Z}_{s,d})_L^2 \left(1-\frac{5}{8}\frac{m_s^2}{M_{Z_D}^2}\right)\,.
\label{eq:delMK-NP}
\eeq
 The experimentally measured value  is given by~\cite{FlavourLatticeAveragingGroup:2019iem}
\beq
\Delta M_K = 3.484(6) \times 10^{-12}~\rm{MeV}\,,
\eeq
which is in accord with the SM expectation.
We find that for $M_{Z_D} = 10$~MeV and $\eps_Z = 0.001$, $\Delta M_K^{NP} = -4.64 \times 10^{-15}$~MeV which is consistent with measurement. Hence, $\Delta M_K$ does not impose a strong constraint on the model.

\subsection{Radiative $K^+ \to \mu^+ \nu_\mu Z_D$ decays}
The three body decay $K^+ \to \mu^+ \nu_\mu Z_D$ can be considered as a radiative correction to the standard $K^+ \to \mu^+ \nu_\mu$ decay in which $Z_D$ is radiated off the muon leg and decays invisibly. Such dark emissions are likely when the dark boson mass is below $2m_\mu$ and is produced on shell.

For a general interaction of the dark boson with the muon,
\beq
\mathcal{L}_D^\mu = \bar{\mu} (g_V \gamma_\al + g_A \gamma_\al \gamma^5) \mu Z_D^\al\,,
\label{eq:radiative-decay-gen}
\eeq 
the amplitude squared of the radiative decay process $K^+ \to \mu^+ \nu_\mu Z_D$ is
\begin{align}
\sum_{spins} |\mathcal{M}|^2 = \frac{G_F^2 f_K^2 |V_{us}|^2}{(Q^2-m_\mu^2)^2}\frac{1}{M_{Z_D}^2} & \left( g_L^2 Q^4 (2 E_\mu m_K (m_\mu^2 + M_{Z_D}^2 - Q^2) - 2 E_D m_K M_{Z_D}^2 \right. \nn \\& \left. - m_K^2 m_\mu^2 + m_K^2 Q^2 - m_\mu^4 + m_\mu^2 Q^2 + 2 M_{Z_D}^4 \right. \nn \\& \left. - m_\mu^2 M_{Z_D}^2 - M_{Z_D}^2 Q^2) + 6 g_L g_R m_\mu^2 M_{Z_D}^2 Q^2 (Q^2 - m_K^2) \right. \nn \\& \left. + g_R^2 m_\mu^2 (-2 E_\mu m_K^3(m_\mu^2 + M_{Z_D}^2 - Q^2)  \right. \nn \\& \left.+ 2 E_D (E_\nu M_{Z_D}^2 (m_K^2 -Q^2) + m_K^3(Q^2 - m_\mu^2) \right. \nn \\& \left. + m_K Q^2 (m_\mu^2 + M_{Z_D}^2 - Q^2)) + M_K^4 m_\mu^2 + m_K^4 M_{Z_D}^2 \right. \nn \\& \left. - M_K^4 Q^2 + m_K^2 m_\mu^4 - m_K^2 m_\mu^2 Q^2  - 3 m_K^2 M_{Z_D}^4  \right. \nn \\& \left. + 2 m_K^2 m_\mu^2 M_{Z_D}^2 + M_{Z_D}^4 Q^2 - m_\mu^2 M_{Z_D}^2 Q^2)\right)\,,
\label{eq:ampsq-radiative-decay}
\end{align} 
where $g_{L(R)} = g_V \mp g_A$. For a final state neutrino with energy $E_\nu$ and a muon with energy $E_\mu$, $Q^2 = (k-q)^2 = m_K^2 - 2 m_K E_\nu$ is the momentum transfer and $E_D$ is the energy of the emitted dark boson. The three-body decay width is given by
\beq 
\Gamma (K \to \mu \nu Z_D) = \frac{1}{64 \pi^3 m_K} \int \sum_{spins} |\mathcal{M}|^2 dE_\mu dE_\nu\,,
\eeq 
with integration limits, 
\beq
E_\mu^{min} = m_\mu\,, ~~~~~ E_\mu^{max} = \frac{m_K^2 + m_\mu^2 - M_{Z_D}^2}{2 m_K}\,,
\eeq 
\beq
E_\nu^{min} = \frac{m_K^2 + m_\mu^2 -M_{Z_D}^2 - 2m_K E_\mu}{2\left(m_K - E_\mu + \sqrt{E_\mu^2 - m_\mu^2}\right)}\,,~~~~ E_\nu^{max} = \frac{m_K^2 + m_\mu^2 -M_{Z_D}^2 - 2m_K E_\mu}{2\left(m_K - E_\mu - \sqrt{E_\mu^2 - m_\mu^2}\right)}\,.
\eeq 

\begin{figure}[t]
    \centering
    \includegraphics[scale=0.655]{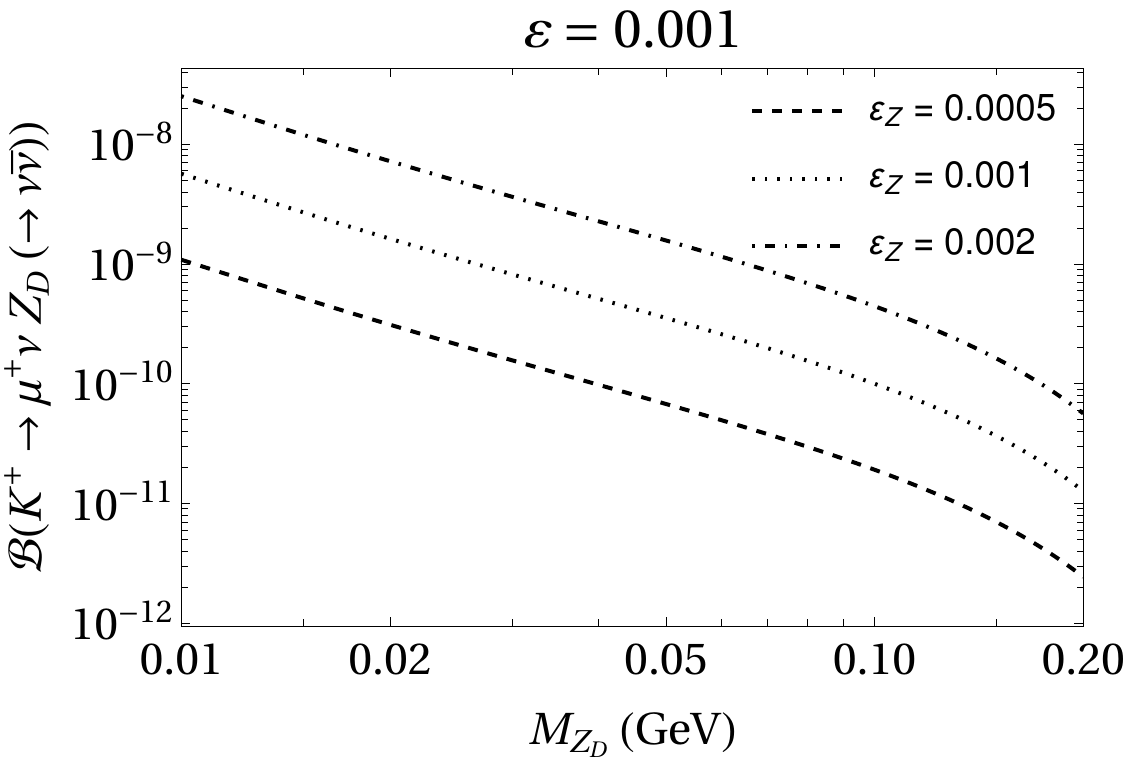}
    \includegraphics[scale=0.645]{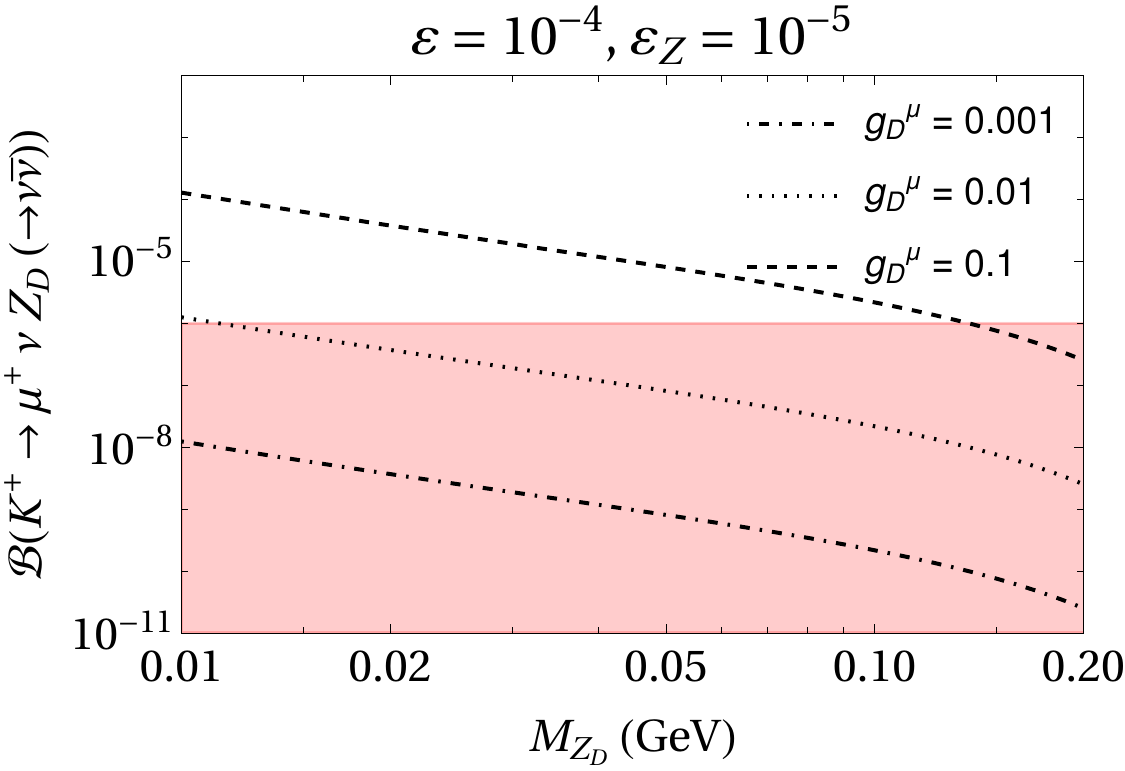}
    \caption{Dependence of the $K^+ \to \mu^+ + invisible$ branching fraction on the mixing parameters in Case A (left) and on the direct coupling $g_{D}^\mu$ in Case B (right). The red shaded region shows the 90\% CL upper limit on the branching fraction in Eq.~\eqref{eq:KtoMuNuInv}.}
    \label{fig:KmunuX}
\end{figure}

The emitted dark boson can subsequently decay to a pair of neutrinos and modify the missing energy spectrum of $K^+ \to \mu^+ \nu$. The NA62 experiment has set the following 
90\%~CL upper limit from searches for kaon decays to single muon final states~\cite{NA62:2021bji}:
\beq
\cB (K^+ \to \mu^+ \nu \nu \bar{\nu}) < 1.0 \times 10^{-6}\,,
\label{eq:KtoMuNuInv}
\eeq
for a signal acceptance of $A_{\mu \nu \nu \nu} = 0.103$. We use this measurement to constrain  the parameter space.
However, since only a dark Z can decay to a pair of neutrinos, the bound does not apply to a dark photon. For Case A, the emission of the dark boson from the muon leg can occur with coupling $g_V = -e\eps + \frac{g}{4c_W}\eps_Z (-1+4s_W^2)$ and $g_A = -\frac{g}{4c_W}\eps_Z$.  The rate is mixing suppressed for this scenario and even for very light masses and $\eps, \eps_Z \sim 0.001$, the branching fraction stays well within the upper limit of Eq.~\eqref{eq:KtoMuNuInv}, as shown in the left panel of Fig.~\ref{fig:KmunuX}.
On the other hand, for Cases B and C, $g_V = -e\eps + \frac{g}{4c_W}\eps_Z (-1+4s_W^2) + g_{D}^\mu$. For small mixings, $g_D^\mu$ provides the dominant contribution and can be tightly constrained. From the right panel of Fig.~\ref{fig:KmunuX}, it can be seen that for Case B, $g_D^\mu$ must be less than 0.01 to satisfy Eq.~\eqref{eq:KtoMuNuInv}.

\begin{figure}[t]
	\centering
	\includegraphics[scale=0.65]{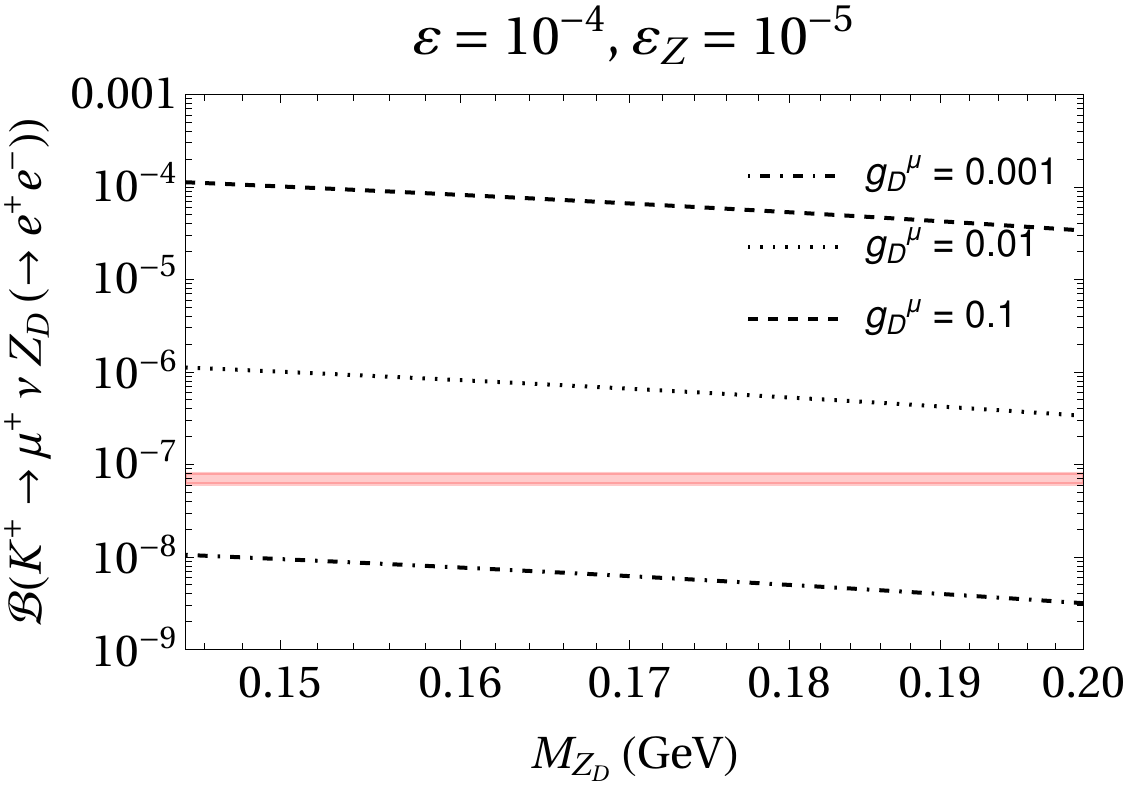}
	\caption{Dependence of the $K^+ \to \mu^+ \nu e^+ e^-$ branching fraction on $g_{D}^\mu$ in Case B. The red shaded region shows the $3\sigma$ interval of the branching fraction in Eq.~\eqref{eq:KtoMuNuee}.}
    \label{fig:KmunuXee}
\end{figure}

The dark $Z$ can also decay to $e^+e^-$ instead of neutrinos. The rate for $K^+ \to \mu^+ \nu e^+ e^-$ has been calculated in the SM and measured in experiments. In the SM, this decay takes place through the emission of a photon from the muon or the kaon (inner bremsstrahlung). The experimentally measured value of the branching fraction of $K^+ \to \mu^+ \nu_\mu e^+ e^-$ is~\cite{ParticleDataGroup:2022pth,Poblaguev:2002ug}
\beq
\mathcal{B}(K^+ \to \mu^+ \nu_\mu e^+ e^-) = (7.06 \pm 0.31) \times 10^{-8}\,,
\label{eq:KtoMuNuee}
\eeq
for the invariant electron-positron mass $m_{ee} > 145$~MeV. In our model, this decay occurs via $K \to \mu \nu Z_D (\to e^+ e^-)$ and the above constraint applies for $M_{Z_D} > 145$ MeV. We find that $\eps, \eps_Z \sim 0.001$ is allowed by the data. For Case B, the dependence of the $K^+ \to \mu^+ \nu e^+ e^-$ branching fraction on $g_{D}^\mu$ is shown in Fig.~\ref{fig:KmunuXee}. Clearly, $g_D^\mu \gsim 0.001$ produces an enhancement in the branching fraction for $145 < M_{Z_D} < 200$~MeV if $\eps_Z < \eps$ and $Z_D$ decays primarily to $e^+ e^-$ below the muon threshold. This constraint does not apply for Case C as $Z_D$ does not couple to electrons.

\subsection{Radiative $\pi^+ \to \mu^+ \nu_\mu Z_D$ decays}
\label{sec:PiMuNu}
\begin{figure}
    \centering
    \includegraphics[scale=0.65]{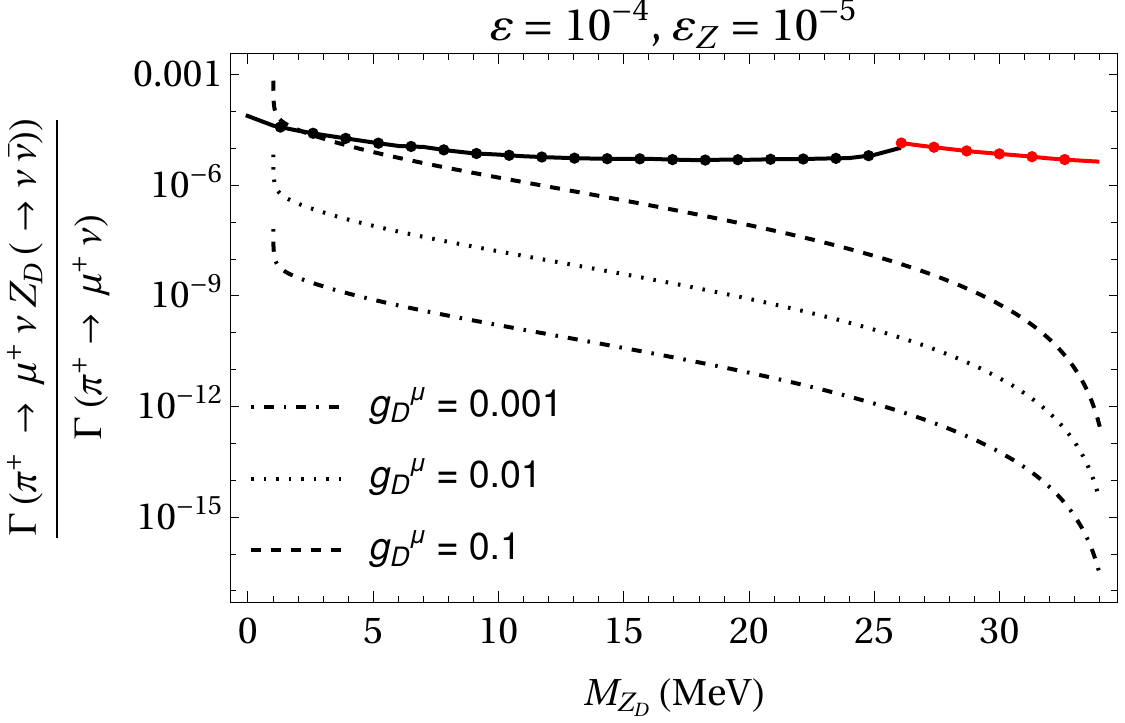}
    \includegraphics[scale=0.65]{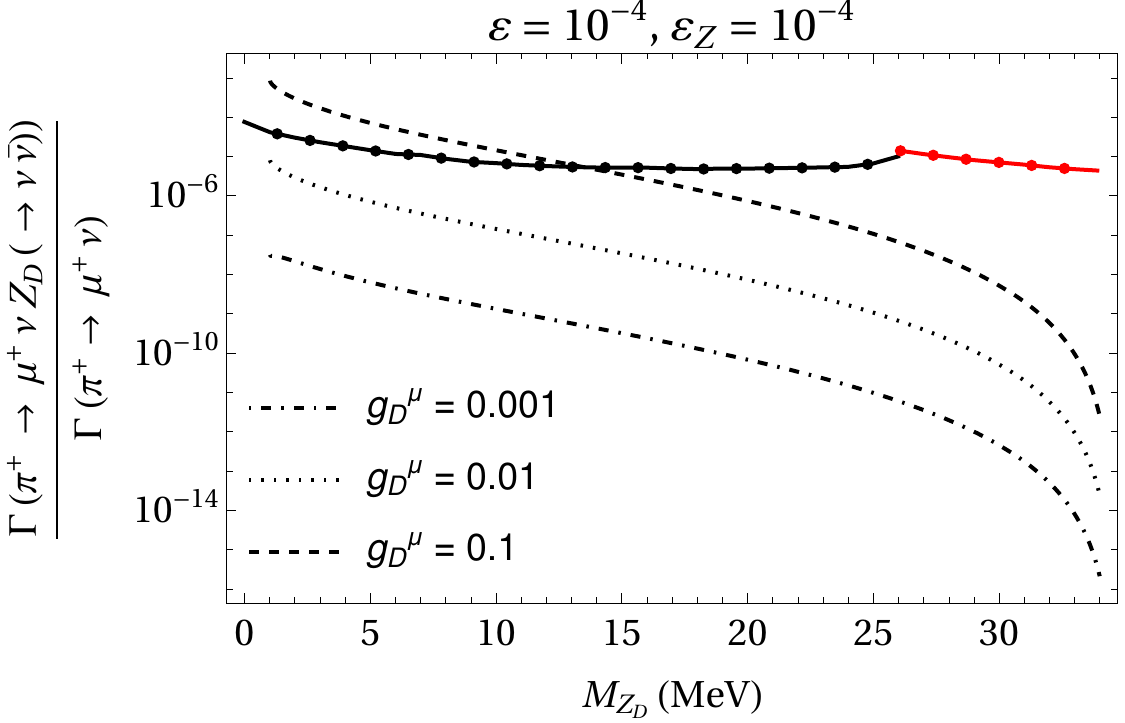}
    \caption{Dependence of $R^{\pi \mu\nu X} = \Gamma(\pi^+ \to \mu^+ \nu_\mu Z_D (\to \nu \bar{\nu}))/\Gamma(\pi^+ \to \mu^+ \nu_\mu)$ on $g_D^\mu$ for $\eps_Z = 10^{-5}$ (left) and $10^{-4}$ (right). The black and red solid curves punctuated with points show the 90\%~CL upper limit from PIENU for the muon kinetic energy ranges, $T_\mu > 1.2$~MeV and $T_\mu < 1.2$~MeV, respectively.  }
    \label{fig:PiMuNuX}
\end{figure}
Similar to radiative kaon decay, $\pi^+ \to \mu^+ \nu Z_D$ can be enhanced for $0 < M_{Z_D} < m_\pi - m_\mu$. The PIENU experiment has presented an upper bound on the ratio $R^{\pi \mu\nu X} = \Gamma(\pi^+ \to \mu^+ \nu_\mu X)/\Gamma(\pi^+ \to \mu^+ \nu_\mu)$ in the range $0 < M_{X} < 33.9$~MeVwhere $X$ decays invisibly~\cite{PIENU:2021clt}. We utilize this bound to constrain our model parameters.

The decay rate is mixing suppressed in Case A and $\sim (g_D^\mu)^2$ for Cases B and C. For the general interaction in Eq.~\eqref{eq:radiative-decay-gen}, the amplitude squared of 
$\pi^+ \to \mu^+ \nu_\mu Z_D$ is given by Eq.~\eqref{eq:ampsq-radiative-decay} with the substitutions $f_K \to f_\pi$, $V_{us} \to V_{ud}$ and $m_K \to m_\pi$. In Fig.~\ref{fig:PiMuNuX}, we show the dependence of $R^{\pi \mu\nu X}$ on $g_D^\mu$. The 90\%~CL upper limit set by PIENU for the muon kinetic energy ranges, $T_\mu > 1.2$~MeV and $T_\mu < 1.2$~MeV, is shown by the solid black and red curves, respectively. We find that $g_{D}^\mu < 0.1$ is allowed for the $M_{Z_D} < 34$~MeV if the dark $Z$ does not primarily decay to neutrinos i.e., if $\eps_Z  \ll \eps$. However, if $\eps_Z \gsim \eps$, $\cB(Z_D \to \nu \bar{\nu})$ is large (unity, if $Z_D$ does not decay to electrons) and $g_D^\mu \geq 0.1$ is not allowed for $M_{Z_D} < 15$~MeV.

\subsection{Atomic parity violation}
\label{sec:APV}
Since the dark $Z$ can couple to first generation SM fermions through mixing, the model parameters are subject to strong constraints from atomic parity violating observables. The interaction of the dark boson with the electromagnetic and weak currents of the SM in Eqs.~\eqref{eq:lag1} and~\eqref{eq:lag2} modify the weak charge $Q_W$ of the proton and nuclei such as cesium ($C_s$), which are measured to be~\cite{Qweak:2018tjf, PDG2022, Wood:1997zq}
\beq
Q_W^{p,exp} = 0.0719(45)\,,\quad Q_W^{^{133}C_s, exp} = -72.82(42)\,.
\label{eq:APV-obs}
\eeq 
The weak charges of the proton and $^{133}C_s$ nuclei are given by~\cite{Cadeddu:2021dqx}
\beq
Q_W^{p,Z_D} = - 2\rho_d g_{AV}^{ep}(\kappa_d \sin^2 \theta_W)\left(1- \frac{\alpha}{2\pi}\right)\,, 
\eeq 
and
\beq
Q_W^{^{133}C_s,Z_D} = - 2\rho_d \left[Z_{C_s} (g_{AV}^{ep}(\kappa_d \sin^2 \theta_W + 0.00005) + N_{C_s}(g_{AV}^{en} + 0.00006) \right]\left(1- \frac{\alpha}{2\pi}\right)\,,
\eeq
including radiative corrections. Here, $g_{AV}^{ep (en)}$ is the effective electron-proton (electron-neutron) coupling and is equal to $-1/2 + 2 \sin^2 \theta_W\ (1/2)$ in the SM at tree level.  The Fermi constant $G_F$ and the weak mixing angle $\theta_W$ are modified by the dark $Z$ interaction to
\beq
G_F \to \rho_d G_F\,,\quad \sin^2\theta_W(Q^2) \to \kappa_d \sin^2\theta_W(Q^2)\,,
\eeq
where
\beq
\rho_d = 1 + \left(\frac{m_Z}{M_{Z_D}}\eps_Z + \frac{M_{Z_D}}{m_Z} \eps \tan \theta_W \right)^2 f\left(\frac{Q^2}{M_{Z_D}^2}\right)\,,
\eeq
and
\beq
\kappa_d  = 1 - \eps\left(\frac{m_Z}{M_{Z_D}}\eps_Z + \frac{M_{Z_D}}{m_Z} \eps \tan \theta_W \right)\frac{m_Z}{M_{Z_D}}\cot \theta_W f\left(\frac{Q^2}{M_{Z_D}^2}\right)\,.
\eeq
For the proton, $f(Q^2/M_{Z_D}^2) = M_{Z_D}^2/(Q^2 + M_{Z_D}^2)$ where $Q^2$ is the momentum transfer. For cesium, $f$ is a constant that depends on $M_{Z_D}$. For example, $f \simeq 0.5$ for $M_{Z_D} \approx 2.4$~MeV while $f \simeq 1$ for $M_{Z_D} \approx 100$~MeV. We follow Ref.~\cite{Cadeddu:2021dqx} to infer $f(Q^2/M_{Z_D}^2)$ for intermediate values of $M_{Z_D}$. The number of neutrons and protons in the cesium nucleus is $N_{C_s} = 78$ and $Z_{C_s} = 55$, respectively.

\begin{figure}[t]
    \centering
    \includegraphics[scale=0.75]{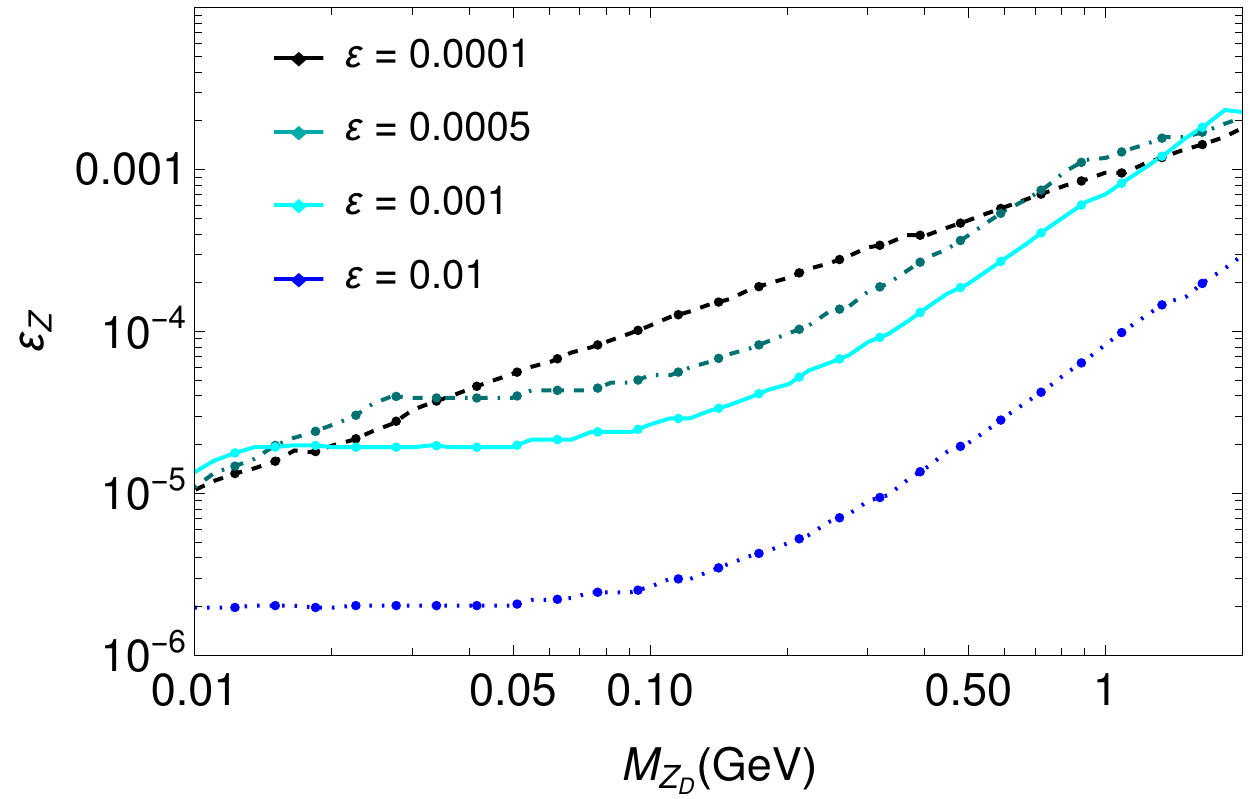}
    \caption{The $3\sigma$ CL upper bound on $\eps_Z$ from measurements of the proton and cesium weak charges in atomic parity violation experiments.}
    \label{fig:APV}
\end{figure}

In Fig.~\ref{fig:APV}, we plot the $3\sigma$ CL upper bound from APV on $\eps_Z$  for different values of $\eps$. By and large, for larger values of $\eps$, 
$\eps_Z$ is more constrained. 
Among the constraints discussed so far, APV places the strongest constraint on $\eps_Z$ in the few MeV-GeV mass range. The coupling $g_{D}^\mu$ which appears in Case B is unconstrained by APV. Again, because of our fine-tuned choice of $g_{D}^e$ to cancel 
the $Z_D$ coupling to electrons,
Case C is also unconstrained by APV.

\subsection{Neutrino trident and CE$\nu$NS}

\begin{figure}[t]
    \centering
    \includegraphics[scale=0.4]{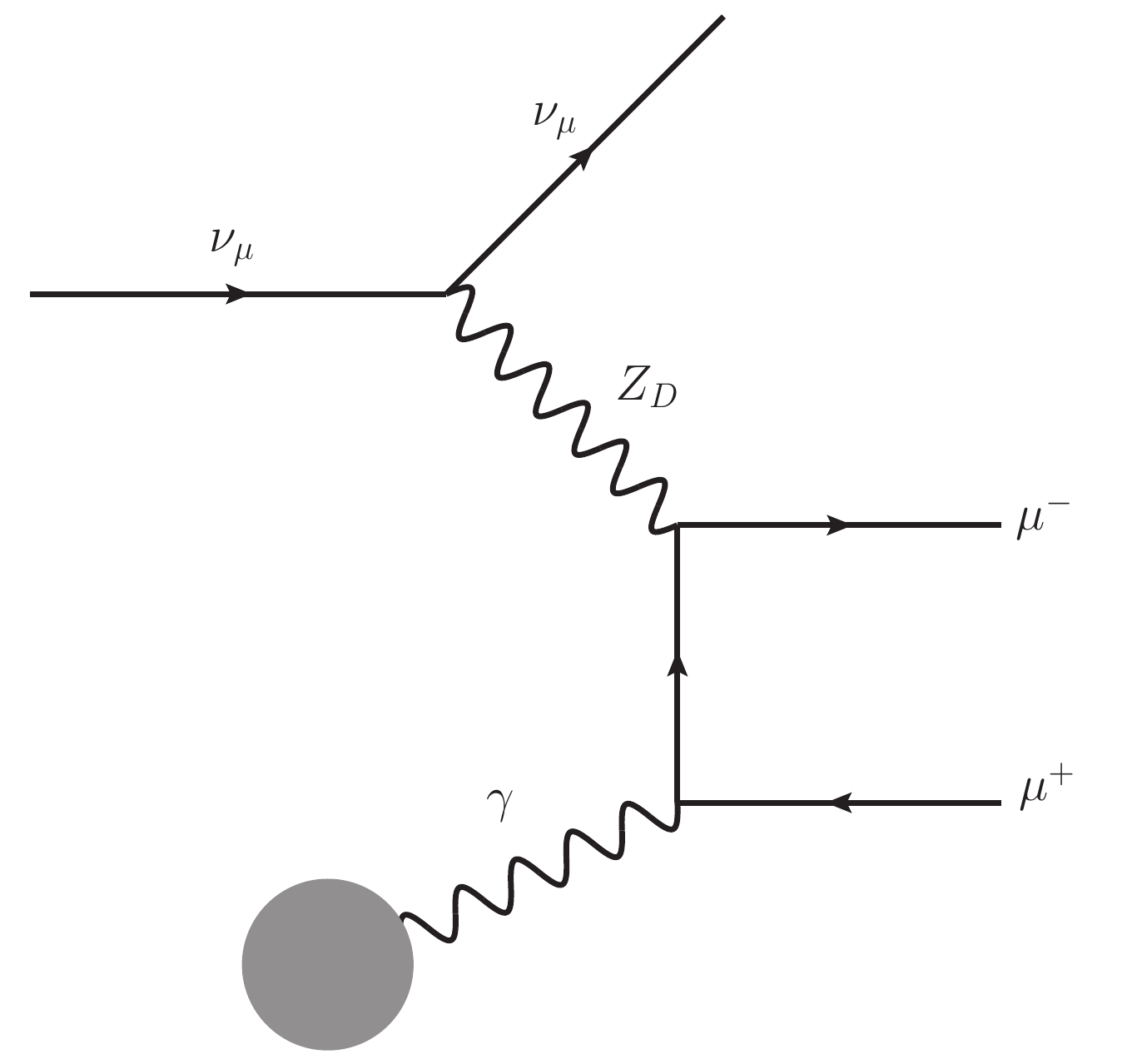}
    \caption{Neutrino trident mediated by $Z_D$.}
    \label{fig:NTP}
\end{figure}

Muon neutrinos can scatter off a nucleus and produce a pair of muons in a weak process known as neutrino trident production. The muon pair is produced through $Z_D$ exchange  as shown in Fig.~\ref{fig:NTP}. The neutrino trident has been measured in several neutrino beam experiments including CHARM-II \cite{CHARM-II:1990dvf} and CCFR~\cite{CCFR:1991lpl}: 
\beq 
\frac{\sigma(\nu_\mu \to \nu_\mu \mu^+\mu^-)_\text{exp}}{\sigma(\nu_\mu \to \nu_\mu \mu^+\mu^-)_\text{SM}} = 
\begin{cases}
1.58 \pm 0.64  & \text{(CHARM-II)} \\ 
0.82 \pm 0.28  & \text{(CCFR).}
\end{cases}
\label{eq:trident_exp}
\eeq

Nonobservation of any excess puts strong upper limits on the couplings of $Z_D$ with muons. 
In Ref.~\cite{Altmannshofer:2019zhy},  bounds have been placed on the gauge coupling and mass of a $Z^\prime$ based on the $U(1)_{L_\mu -L_\tau}$ symmetry. For Cases B and C, we repurpose the bound in the $(M_{Z^\prime}, g^\prime)$ plane to the $(M_{Z_D}, \eps_Z)$ plane by the scaling $(g^\prime)^2 \to g\eps_Z g_{D}^{\mu}/4c_W$, where the $\eps_Z$ suppression comes from the $Z_D \nu \nu$ vertex. Since the bound on $g^\prime$ was derived by assuming vector-like interactions of the lepton, it cannot be applied to Case A which has both vector and axial-vector interactions with the muon. However, we anticipate that a CCFR bound on Case A will not be as stringent as the APV bound.

The COHERENT experiment first observed coherent elastic neutrino-nucleus scattering (CE$\nu$NS) in a CsI detector~\cite{COHERENT:2017ipa} and subsequently in a liquid Argon detector~\cite{COHERENT:2020iec,COHERENT:2020ybo}.  The data have been used to set upper limits on gauged $U(1)_X$ model parameters~\cite{Han:2019zkz, Banerjee:2021laz, AtzoriCorona:2022moj}. In our model,  CE$\nu$NS can occur via the exchange of $Z_D$ and so COHERENT data constrain $\eps_Z$ for a given $Z_D$ mass. 
We rescale bound from Ref.~\cite{Han:2019zkz} according to $g^\prime \to \eps_Z \frac{g}{c_W} g_V^\nu g_V^{q}$ with $g_V^\nu = 1/2,~g_V^u = \frac{1}{2} - \frac{4}{3}s_W^2,~g_V^d = -\frac{1}{2} + \frac{2}{3}s_W^2$. Bounds on $\eps_Z$ turn out to be much weaker than on a direct gauge coupling $g^\prime$. 
For example, we find $\eps_Z \lsim 0.0005$ for $M_{Z_D} \lsim 10$~MeV.

\subsection{Collider and other bounds}
\label{sec:collider-bounds}
It is possible for $Z_D$ to be produced on- or off-shell from $Z$ decay. An example is $Z \to \ell \ell Z_D$ where $Z_D$ is radiated from one of the lepton legs. $Z_D$ may then decay to a pair of leptons or jets, which lead to $Z \to 4 \ell$ ($\ell = e, \mu$), final states that have been searched for at ATLAS~\cite{ATLAS:2014jlg, ATLAS:2022wrd} and CMS~\cite{CMS:2012bw, CMS:2018yxg}. The results are consistent with the SM. The invariant $4\ell$ mass peaks at the $Z$ mass and places a bound for $M_{Z_D} \gsim 5$~GeV. For our model, the $Z_D$ couplings to leptons are suppressed by the mixing parameters making the decay rate for
$Z_D \to 4 \ell$ small. Also, the emitted dark boson is expected to be soft and therefore will not produce energetic leptons in the final state. Hence, the bounds from this decay do not impact our model. The same applies to NP searches in the jets + missing energy final state.

Recently, Belle II searched for an invisibly decaying $Z^\prime$ boson in $e^+e^- \to \mu^+\mu^- +$ missing energy channel~\cite{Belle-II:2019qfb}. The search puts an upper limit on the $Z^\prime$ coupling  to muons to be between $0.05 - 1$ for $M_{Z'} < 6$~GeV under the assumption that $Z^\prime$ decays only to invisible particles. The bound becomes weaker for $\cB(Z^\prime \to invisible) < 1$. Consequently, this limit is not as competitive as the low energy bounds discussed previously.

The dark boson can contribute to the leptonic decay width of the $W$ boson, which is measured very precisely. Additional contributions to $\Gamma(W \to \ell + invisible)$ arising from $\Gamma(W \to \ell \nu Z_D, Z_D\to \nu \bar{\nu})$ can be significant due to the enhancement from the longitudinal polarization of the $Z_D$. In the limit $m_\ell, M_{Z_D} \ll M_W$, if $Z_D$ decays mostly invisibly, the three body leptonic decay width is given by~\cite{Darme:2021qzw}
\beq
\Gamma(W \to \ell \nu Z_D) \simeq \frac{(C_{\ell \ell Z_D} - \widetilde{C}_{\ell \ell Z_D})^2 G_F M_W^5}{512\sqrt{2}\pi^3 M_{Z_D}^2}\,,
\label{eq:WtolnuZD}
\eeq
where $C_{\ell \ell Z_D},\, \widetilde{C}_{\ell \ell Z_D}$ are the effective vector and axial-vector $Z_D \ell \ell$ couplings, respectively. Fo a dark photon, $C_{\ell \ell Z_D} = e \eps$ and $\widetilde{C}_{\ell \ell Z_D} =0$, and for a dark $Z$, $C_{\ell \ell Z_D} = \frac{g}{\cos \theta_W}\eps_Z g_V^\ell$ and $\widetilde{C}_{\ell \ell Z_D} =\frac{g}{\cos \theta_W}\eps_Z g_A^\ell$. The new contribution to the $W \to \ell + invisible$ decay is then given by
\beq 
\Gamma(W \to \ell + invisible) = \Gamma(W \to \ell \nu Z_D) \cdot \cB(Z_D \to \nu \bar{\nu})\,.
\eeq  
We conservatively require the new contribution to lie within the uncertainty of the measured decay width of the $W$ boson. For Case A, if the dark boson mass is below the dimuon threshold, $\cB (Z_D \to \nu \bar{\nu}) \sim 70\%$ for $\eps_Z \sim \mathcal{O}(10^{-3})$. Then, since $\Gamma_W = 2.085 \pm 0.042$~GeV~\cite{PDG2022}, the $2\sigma$ bound on the mixing parameters is given by
\beq 
 \eps_Z \lsim 0.08\left(\frac{M_{Z_D}}{100 \text{ MeV}} \right)\,,
\label{eq:W-width-bound-1}
\eeq
which is not competitive with our previous  bounds. The bound gets more restrictive if there is a direct coupling of the dark $Z$ to muons, as we discuss later.

The LHCb collaboration has performed searches for dark photons in dimuon samples~\cite{LHCb:2019vmc} and put stringent bounds on the mixing parameter.  It is found that $\eps \lsim 10^{-4}$ for $M_{Z_D} \sim 200$ MeV and $\eps \lsim 0.0005$ for $M_{Z_D} \sim 2$~GeV at the 90\% CL. This bound directly applies to Case A, and for Cases B and C, we recast the bound on $\eps$ to the coupling $g_{D}^\mu$ by appropriately scaling the effective $pp \to Z_D \to \mu \mu$ interaction strength from $(e\eps)^2 \to (e\eps)g_D^\mu$. This rules out $g_D^\mu > 10^{-3}$ for $M_{Z_D} > 210$~MeV.

\section{Parameter fits}

\begin{table}[t]
\centering
\footnotesize
\renewcommand{\arraystretch}{1.2}
\resizebox{\columnwidth}{!}{\begin{tabular}{|c|c|c|c|c|}
\hline
\textbf{Decay} & \textbf{Ref.} & $\mathbf{q^2}$ \textbf{bin (GeV$^2$)}  &  \textbf{Measurement} & \textbf{SM expectation} \\
\hline
\multirow{4}{*}{$\frac{d\mathcal{B}}{dq^2}(B^0 \to K^{*0} \mu^+ \mu^-) \times 10^{8}$} & \multirow{4}{*}{\cite{LHCb:2016ykl}} & $0.1-0.98$ & $11.06^{+0.67}_{-0.73}\pm 0.29 \pm 0.69$ & $10.60 \pm 1.54$ \\
& & $1.1-2.5$ & $3.26^{+0.32}_{-0.31}\pm 0.10 \pm 0.22$ & $4.66\pm 0.74$ \\
& & $2.5-4.0$ & $3.34^{+0.31}_{-0.33}\pm 0.09 \pm 0.23$ & $4.49\pm 0.70$ \\
& & $4.0-6.0$ & $3.54^{+0.27}_{-0.26}\pm 0.09 \pm 0.24$ & $5.02\pm 0.75$ \\
\cline{1-5}
\multirow{3}{*}{$\frac{d\mathcal{B}}{dq^2}(B^+ \to K^{*+} \mu^+ \mu^-) \times 10^{8}$} & \multirow{3}{*}{\cite{LHCb:2014cxe}} & $0.1-2.0$ & $5.92^{+1.44}_{-1.30}\pm 0.40$ & $7.97\pm 1.15$ \\
& &  $2.0-4.0$ & $5.59^{+1.59}_{-1.44}\pm 0.38$ & $4.87\pm 0.76$ \\
& & $4.0-6.0$ & $2.49^{+1.10}_{-0.96}\pm 0.17$ & $5.43\pm 0.74$ \\
\cline{1-5}
\multirow{6}{*}{$\frac{d\mathcal{B}}{dq^2}(B^+ \to K^{+} \mu^+ \mu^-) \times 10^{8}$} & \multirow{6}{*}{\cite{LHCb:2014cxe}} & $0.1-0.98$ & $3.32\pm 0.18 \pm 0.17$ & $3.53 \pm 0.64 $ \\
& & $1.1-2.0$ & $2.33\pm 0.15 \pm 0.12$ & $3.53 \pm 0.58$ \\
& & $2.0-3.0$ & $2.82\pm 0.16 \pm 0.14$ & $3.51 \pm 0.52$ \\
& & $3.0-4.0$ & $2.54\pm 0.15 \pm 0.13$ & $3.50 \pm 0.63$ \\
& & $4.0-5.0$ & $2.21\pm 0.14 \pm 0.11$ & $3.47 \pm 0.60$ \\
& & $5.0-6.0$ & $2.31\pm 0.14 \pm 0.12$ & $3.45 \pm 0.53$ \\
\cline{1-5}
\multirow{3}{*}{$\frac{d\mathcal{B}}{dq^2}(B^0 \to K^{0} \mu^+ \mu^-) \times 10^{8}$} & \multirow{3}{*}{\cite{LHCb:2014cxe}} & $0.1-2.0$ & $1.22^{+0.59}_{-0.52} \pm 0.06$ & $3.28 \pm 0.52$ \\
& & $2.0-4.0$ & $1.87^{+0.55}_{-0.49} \pm 0.09$ & $3.25 \pm 0.56$ \\
& & $4.0-6.0$ & $1.73^{+0.53}_{-0.48} \pm 0.09$ & $3.21 \pm 0.54$ \\
\cline{1-5}
\multirow{4}{*}{$\frac{d\mathcal{B}}{dq^2}(B_s^0 \to \phi \mu^+ \mu^-) \times 10^{8}$} & \multirow{4}{*}{\cite{LHCb:2021zwz}} & $0.1-0.98$ & $7.74 \pm 0.53 \pm 0.12 \pm 0.37$ & $11.31 \pm 1.34$ \\
& & $1.1-2.5$ & $3.15\pm 0.29 \pm 0.07 \pm 0.15$ & $5.44 \pm 0.61$ \\
& & $2.5-4.0$ & $2.34 \pm 0.26 \pm 0.05 \pm 0.11$ & $5.14 \pm 0.73$ \\
& & $4.0-6.0$ & $3.11 \pm 0.24\pm 0.06 \pm 0.15$ & $5.50 \pm 0.69$ \\
\cline{1-5}
\multirow{2}{*}{$\mathcal{B}(B^+ \to K^{+} e^+e^-) \times 10^{8}$} & \multirow{2}{*}{\cite{BELLE:2019xld}} & $0.1-4.0$ & $18.0^{+3.3}_{-3.0} \pm 0.5$ & $13.73 \pm 1.88$ \\
& & $4.0-8.12$ & $9.6^{+2.4}_{-2.2} \pm 0.3$ & $14.11 \pm 1.88$ \\
\cline{1-5}
$\mathcal{B}(B^0 \to K^{*0}e^+ e^-) \times 10^{7}$ & \cite{LHCb:2013pra} &$0.03^2-1.0^2$ & $3.1^{+0.9+0.2}_{-0.8-0.3} \pm 0.2$ & $2.56 \pm 0.44$ \\
\cline{1-5}
$\mathcal{B}(B \to X_s \mu^+ \mu^-) \times 10^{6}$ & \cite{BaBar:2013qry} & $1.0-6.0$ & $0.66^{+0.82+0.30}_{-0.76-0.24} \pm 0.07$ & $1.67 \pm 0.15$ \\
$\mathcal{B}(B \to X_s e^+ e^-) \times 10^{6}$ & \cite{BaBar:2013qry} & $1.0-6.0$ & $1.93^{+0.47+0.21}_{-0.45-0.16} \pm 0.18$ & $1.74 \pm 0.16$ \\
\hline
$\frac{d\mathcal{B}}{dq^2}(B^+ \to K^{+} e^+e^-) \times 10^{9}$ & \cite{LHCb:2022zom} & $1.1-6.0$ & $25.5^{+1.3}_{-1.2}\pm 1.1$ & $34.9\pm 6.2$\\
\hline
$\frac{d\mathcal{B}}{dq^2}(B^0 \to K^{*0} e^+e^-) \times 10^{9}$ & \cite{LHCb:2022zom} & $1.1-6.0$ & $33.3^{+2.7}_{-2.6}\pm 2.2$ & $47.7\pm 7.5$\\
\hline
\end{tabular}}
\caption{Experimental measurements and SM expectations in $q^2$ bins. The SM $\chi^2$ for the fit to all the observables is 93.56, and for just the muon modes it is 84.30.}
\label{tab:data}
\end{table}

We fit the most recent experimental data for the exclusive decays, $B \to K^{(*)} \ell^+ \ell^-$ and $B_s^0 \to \phi \mu^+ \mu^-$, as well as the inclusive decays, $B \to X_s \ell^+ \ell^-$, in different $q^2$ bins as listed in Table~\ref{tab:data}. Note that we only analyze data below the $c \bar{c}$ resonances. We use \texttt{flavio}~\cite{Straub:2018kue} to calculate the SM and NP expectations. We determine the allowed parameter values (denoted X) for each model using
\beq
\chi^2 (X) = \sum_{i=1}^{n} \frac{(\cB_i^{th}(X)-\cB_i^{exp})^2}{\sigma_{i}^2}\,,
\eeq
where $n$ is the number of measurements, $\cB^{th}$ is the branching fraction, $\cB^{exp}$ is the corresponding experimental value and $\sigma$ is the uncertainty from measurement and theory added in quadrature. For the theoretical uncertainty, we take the SM uncertainty estimated by \texttt{flavio}. To produce two-dimensional allowed regions, we marginalize over the third parameter and define the 1$\sigma$, $2\sigma$ and $3\sigma$ CL regions by  $\chi^2 \leq \chi^2_{\text{min}} + 2.30$, $\chi^2 \leq \chi^2_{\text{min}} + 6.17$, and $\chi^2 \leq \chi^2_{\text{min}} + 11.83$, respectively, where $\chi^2_{\text{min}}$ is the global minimum. We define the pull with respect to the SM as $\sqrt{\chi^2_{SM}-\chi^2}$.

For Cases A and B, we fit all 27 observables in Table~\ref{tab:data}, while for Case C, we exclude the 6 electron mode observables.
$\chi^2_{SM}=93.56$ for the fit to muon and electron mode observables, and $\chi^2_{SM}=84.30$ for the fit to only muon mode observables. 

\subsection{Case A}
\begin{figure}[t]
    \centering
    \includegraphics[scale=0.6]{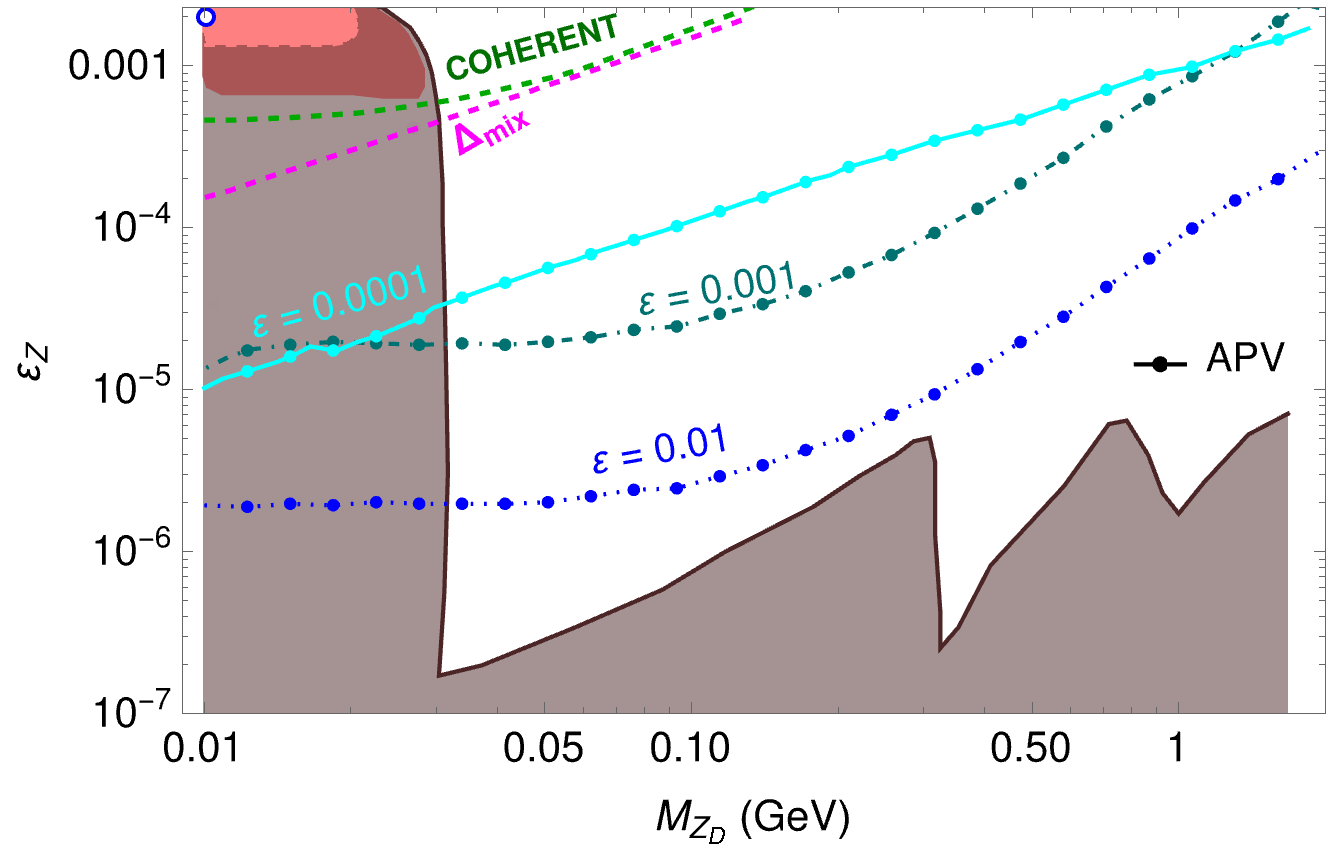}
    \includegraphics[scale=0.6]{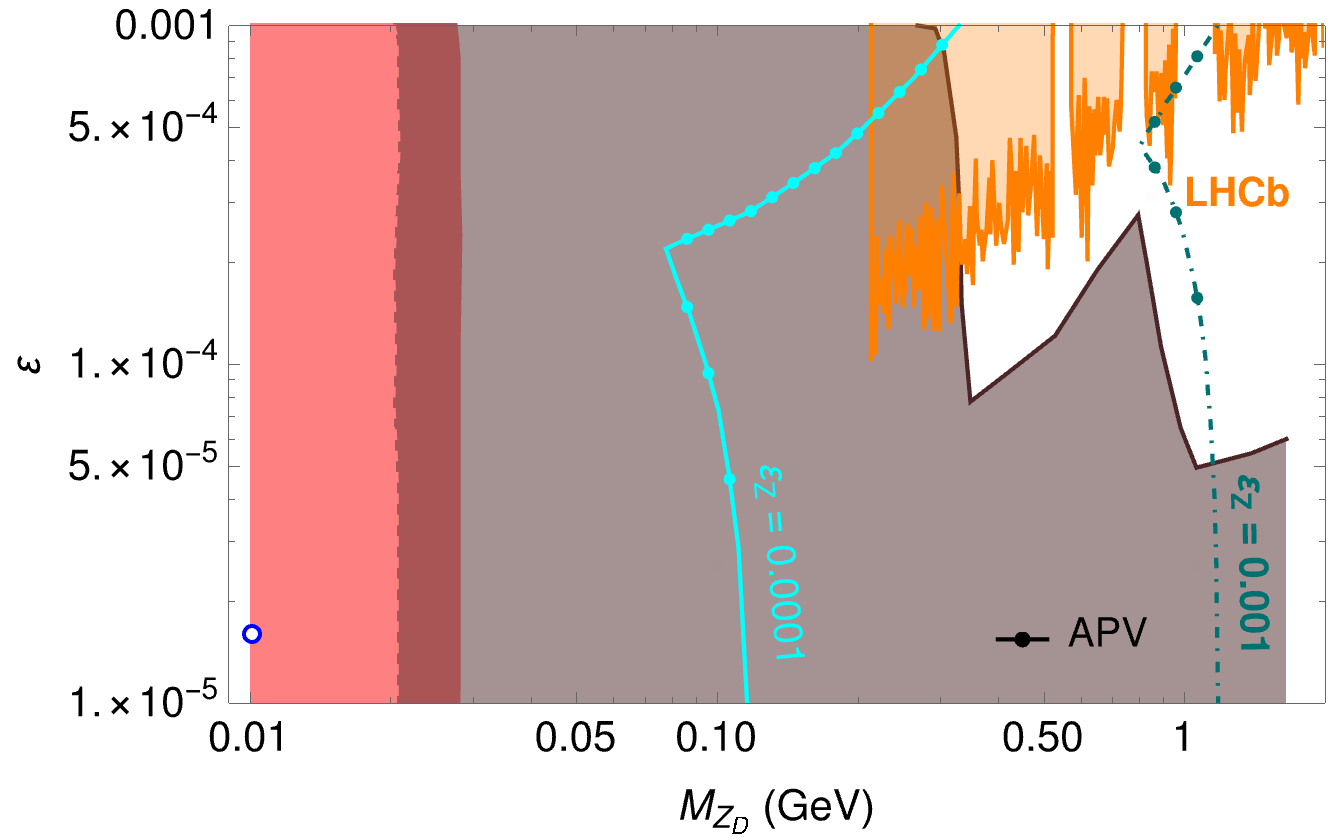}
    \caption{The $1\sigma$ (pink), $2\sigma$ (brown) and 3$\sigma$ (dark brown) regions allowed by the data in Table~\ref{tab:data} for Case A. The best fit point is marked by the blue circle. Top panel: $2\sigma$ upper limit from $B_s^0$-$\overline{B}_s^0$ mixing and $1\sigma$ upper limit from COHERENT neutrino scattering data are shown by the dashed magenta and green curves, respectively. The $3\sigma$ upper limits on $\eps_Z$ from the APV measurements for $\eps = 0.0001, 0.001$ and $0.01$ are shown by the cyan, dark cyan and blue dotted curves, respectively.
    Bottom panel: The orange shaded region is excluded by LHCb dark photon searches at the 90\% CL. The $3\sigma$ upper limits on $\eps$ from APV measurements for $\eps_Z = 0.0001$ and $0.001$ are shown by the cyan and dark cyan curves, respectively; regions to the left of the curves are excluded.}
    \label{fig:CaseA}
\end{figure}

In accordance with the bounds of the previous section, we restrict $\eps \leq 0.001$ and $\eps_Z \leq 0.002$. Our fit to the data shows an improvement with respect to the SM with $\chi^2_{NP} = 85.8$ and pull $=2.79$ at the best fit point,
\beq
M_{Z_D} = 10.07 \text{ MeV}\,, \quad \eps = 1.6\times 10^{-5}\,, \quad \eps_Z = 0.002\,.
\label{eq:CaseA-bf}
\eeq 
In Fig.~\ref{fig:CaseA}, we show the marginalized $1\sigma$ (pink), $2\sigma$ (brown) and $3\sigma$ (dark brown) allowed regions in the ($M_{Z_D}, \eps_Z$) and ($M_{Z_D}, \eps$) planes. The best fit point is marked by a blue circle. 
Above $M_{Z_D} = 30$~MeV, $Z_D$ contributes as a resonance to the $B^0 \to K^{*0} e^+ e^-$ branching fraction in the $[0.03^2,1^2]$~GeV$^2$ bin, and we use the narrow width approximation to obtain the branching fraction.
The sharp drop in the allowed values of $\eps_Z$ above 30~MeV is a consequence of this resonance. The bound relaxes gradually as the mass increases until about $M_{Z_D} \sim 300$~MeV when $Z_D$ starts appearing as a narrow resonance in the  0.1~GeV$^2$ bin for $b \to s \mu \mu$, and causes another sharp drop in $\eps_Z$. Similar drops in the allowed values of $\eps$ occur for the same reason. 
However, since the $\eps$ dependent $b \to s$ loop coefficient is suppressed compared to the $q^2$ independent, $\eps_Z$ dependent monopole contribution, the limits on $\eps$  are less stringent than those on $\eps_Z$. 
The expected drop around $M_{Z_D} = 30$~MeV is not visible because we restrict the plot to $\eps \leq 0.001$.

The model parameters are, however, in conflict with some low energy constraints. In particular, the weak charges measured in APV experiments rule out the entire allowed parameter space for $M_{Z_D} \lsim 30$~MeV at more than $3\sigma$.  
 An important outcome of fitting the binned $b\to s \ell \ell$ data is that for $M_{Z_D} > 30$~MeV, we obtain upper limits on the mixing parameters that are even  stronger than the APV limits. 

Clearly, additional new physics is needed to reconcile all the data.
To weaken the bound on $\eps_Z$ for $M_{Z_D} < 30$~MeV, we consider an additional contribution to the invisible decay width of the dark boson between $10\%$ and $100\%$ of the decay width to neutrinos. However, we find that this has a negligible impact on the fit. The best fit values of the mixing parameters remain the same with a slight improvement in the $\chi^2$. 

\subsection{Case B}
\label{Sec:Case-B}

\begin{figure}[t]
\centering
    \includegraphics[scale=0.7]{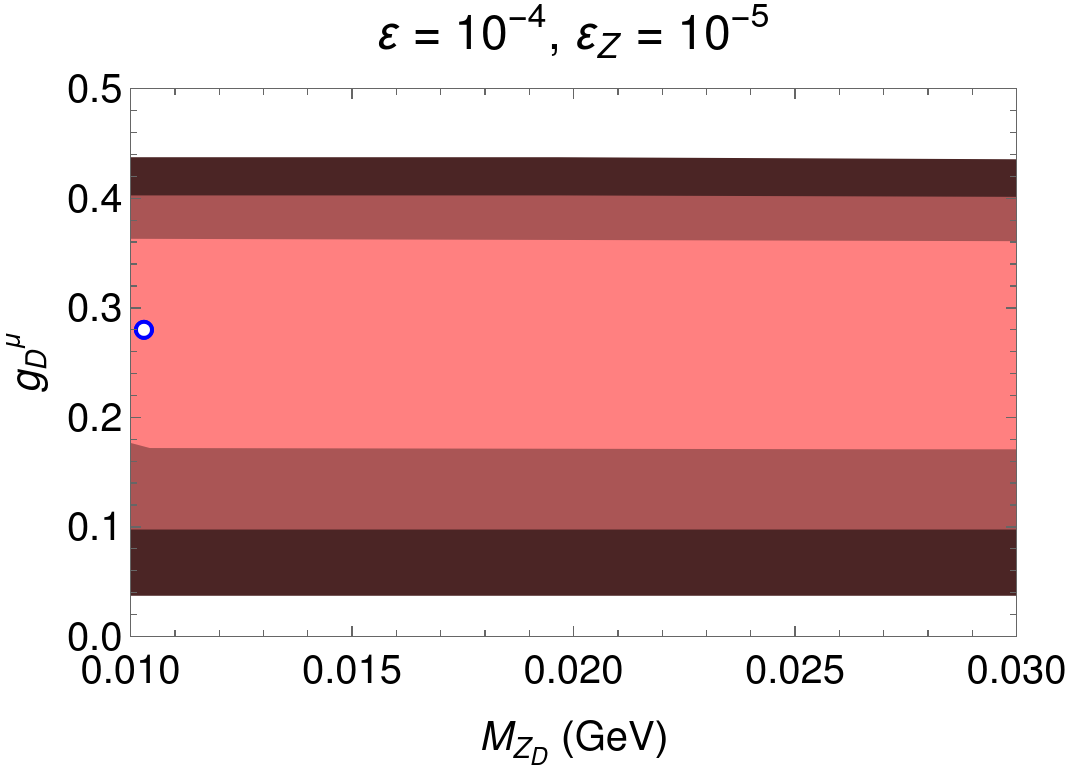}
    \caption{The $1\sigma$, $2\sigma$ and $3\sigma$ allowed regions for Case B with the best fit point marked by a blue circle. However, the entire parameter space is ruled out by
    measurements of $K^+ \to \mu^+ \nu X$,  $X = invisible/e^+ e^-$)  and separately by the $W$ boson width.}
    \label{fig:CaseB}
\end{figure}

The direct interaction of the dark $Z$ with muons results in a contribution to the Wilson coefficient $\mathcal{C}_{9,\mu}$ apart from the mixing induced contribution in Eq.~\eqref{c9}. The hadronic part of the amplitude is still loop induced and proportional to the mixing parameters. In our data analysis, we fix $\eps = 10^{-4}$ and $\eps_Z = 10^{-5}$ to evade stringent constraints from APV and meson mixing. 
We find the best fit point to be  
\beq 
M_{Z_D} = 10.3\ \text{MeV}\,,\quad  g_D^\mu = 0.28\,,
\label{eq:CaseB-bf}
\eeq 
with a $\chi^2_{NP}=75.15$ and pull $=4.29$ from the SM. Case B represents a substantial improvement over Case A. Since there are only two fit parameters in this scenario, no marginalization is needed to obtain the allowed parameter space in Fig.~\ref{fig:CaseB}. 

Although we have chosen the mixing parameters to evade the APV bounds, due to the unsuppressed coupling of $Z_D$ with muons, the $K^+ \to \mu^+ \nu X$ decay rate, where $X = invisible/e^+ e^-$) rules out the entire parameter space in Fig.~\ref{fig:CaseB};  see the right panel Fig.~\ref{fig:KmunuX}. (In Case A, this rate was suppressed by the mixing parameter.)  
The allowed parameter space is separately excluded by measurements of the $W$ boson width because the $Z_D \mu \mu$ interaction contributes to $\Gamma(W \to \mu \nu Z_D)$. As in Section~\ref{sec:collider-bounds}, we require $\Gamma_W = 2.085 \pm 0.042$~GeV~\cite{PDG2022}, and obtain the $2\sigma$ bound,
\beq 
g_D^\mu < 0.022 \left(\frac{M_{Z_D}}{100 \text{ MeV}}\right)\,,
\label{eq:W-width-bound-2}
\eeq
which is in agreement with Ref.~\cite{Darme:2021qzw}. 
We do not show these constraints in Fig.~\ref{fig:CaseB} since the entire region is ruled out.

As a further extension of Case B, consider an additional axial-vector coupling of the dark $Z$ to muons:
\beq 
\mathcal{L}_D^{Z} \supset g_D^\mu \bar{\mu} \gamma_\nu \mu Z_{D}^\nu + g_{DA}^\mu \bar{\mu} \gamma_\nu \gamma_5 \mu Z_{D}^\nu\,.
\label{eq:CaseC}
\eeq
The axial-vector interaction generates a new contribution to the Wilson coefficient $\mathcal{C}_{10,\mu}$. Introduction of an additional parameter improves the quality of the fit and results in $g_{DA}^\mu \sim -g_D^\mu$ which is in accordance with global fits for the scenario, $\mathcal{C}_9 = -\mathcal{C}_{10}$. However, due to additional contributions to the branching fractions of leptonic kaon decays, leptonic $W$ decay, and $\cB(B_s \to \mu \mu)$, we anticipate even more stringent constraints on the parameter space than before. Hence, we do not entertain this possibility any further.

\subsection{Case C}
\begin{figure}[t]
    \centering
    \includegraphics[scale=0.8]{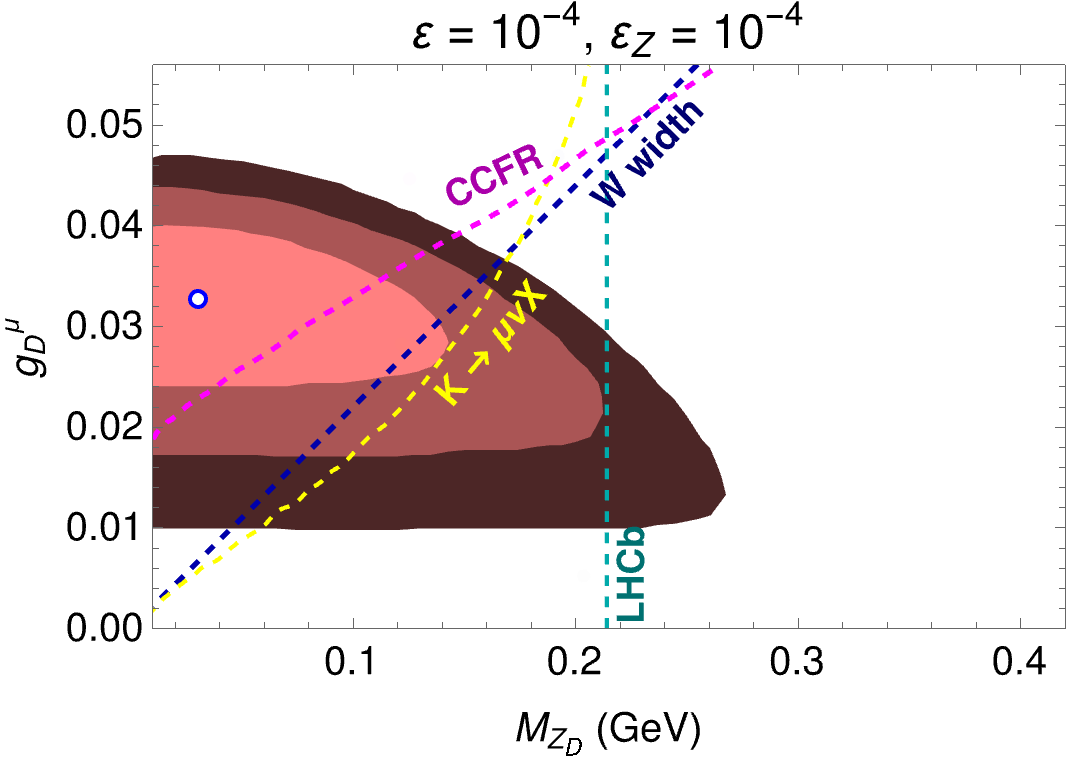}
    \caption{The $1\sigma$, $2\sigma$ and $3\sigma$ allowed regions for Case C with the best fit point marked by a blue circle. 
    Upper limits from neutrino trident production at CCFR (at 95\% CL), $K \to \mu \nu X$ (at 90\% CL) and the $W$ width (at 95\% CL) are shown by the dashed magenta, yellow and dark blue curves, respectively. 
    Dark photon searches at LHCb rule out the region to the right of the vertical dashed dark cyan line at 90\% CL.}
    \label{fig:CaseC}
\end{figure}

In addition to $g_D^\mu$, we now consider a direct coupling $g_{D}^e$  of $Z_D$ to electrons. As mentioned earlier, we assume that $g_{D}^e$ is fine-tuned to cancel the $Z_D$ coupling to electrons via mixing. Then, all observables for the electron mode are well described by the SM. The motivation for this scenario is that it provides an alternate way to bypass the APV constraints and relax bounds on the mixing parameters. We redo the fit to the $b\to s$ data except the 6 electron mode observables in Table~\ref{tab:data}. This reduces the number of $b \to s$ observables from 27 to 21. We set 
$\eps= \eps_Z = 10^{-4}$ to satisfy bounds from $B_s^0-\overline{B}_s^0$ mixing. The best fit point is
\beq
M_{Z_D} = 30.2 \text{ MeV}\,, \quad  g_D^\mu = 0.033\,,
\label{eq:CaseD-bf}
\eeq 
with $\chi^2_{NP} = 56.90$ and pull $= 5.23$ ($\chi^2_{SM} = 84.30$). This is a marked improvement over the previous cases. As expected, $g_D^\mu$ is an order of magnitude smaller than in Case B since $\eps_Z$ is larger by an order of magnitude. The allowed region with exclusions from all applicable bounds is shown in Fig.~\ref{fig:CaseC}.
The most stringent constraints are provided by neutrino trident production at CCFR (dashed magenta, 95\% CL), $K \to \mu+invisible$  (dashed yellow, 90\% CL) and the $W$ width (dashed dark blue, 95\% CL). Dark photon searches at LHCb rule out the region to the right of the vertical dashed dark cyan line at 90\% CL. A region of parameter space bordered by $100 \lsim M_{Z_D} \lsim 200$~MeV and $0.015 \lsim g_D^\mu \lsim 0.035$ explains the binned $b \to s \mu^+ \mu^-$ data at $2\sigma$ CL and is consistent with other bounds.

\section{Muon anomalous magnetic moment}

Measurements of the anomalous magnetic moment of the muon $a_\mu = (g-2)_\mu/2$~\cite{Muong-2:2006rrc,Muong-2:2021ojo} differ from the SM expectation~\cite{Aoyama:2020ynm} by
\beq 
\Delta a_\mu = a_\mu^{exp} - a_\mu^{SM} = 251(59) \times 10^{-11}\,.
\label{eq:delamu}
\eeq
In Case A, $Z_D$ couples to muons via mixing, while in Cases B and C,  $Z_D$ couples directly to muons. The new contribution to $a_\mu$ in the three cases is given by~\cite{Lindner:2016bgg}

\bea
\footnotesize
    a_\mu^{A} &=& \frac{m_\mu^2}{4\pi^2}\int_0^1 \frac{dx}{m_\mu^2 x^2 + M_{Z_D}^2 (1-x)} \left[ (e\eps)^2 x^2(1-x) + \left(\frac{g(-1+4s_W^2)}{4c_W}\eps_Z \right)^2 x^2(1-x) \right. \nonumber \\&& + \left. \left(\frac{g}{4c_W}\eps_Z \right)^2 (x(1-x)(x-4) - 2x^3m_\mu^2/M_{Z_D}^2) \right]\,, \\
    a_\mu^{B,C} &=& \frac{(g_{D}^{\mu} m_\mu)^2}{4\pi^2}\int_0^1 \frac{ x^2(1-x)}{m_\mu^2 x^2 + M_{Z_D}^2 (1-x)} dx\,.
\eea

The FCNC constrained values of $g_D^\mu$ yield values of $a_\mu$ much larger than dictated by Eq.~\eqref{eq:delamu}. Also, the axial-vector contribution to $a_\mu$ in Case A leads to an overall enhancement in the total yield of $a_\mu$.
For the best-fit points, we find $a_\mu^A = -3.45 \times 10^{-7}$, $a_\mu^B = 0.00077$ and $a_\mu^C = 7.38 \times 10^{-6}$.
This suggests  that some other source of new physics is needed to suppress the contribution. One possibility is a dark charged scalar which gives a negative contribution to $a_\mu$.

\section{Summary}
We performed a comprehensive study of the contributions of a dark photon and dark $Z$ to $b \to s \ell^+ \ell^-$ observables.
  We correctly treated the FCNC decay amplitudes via the monopole and dipole operators of the dark bosons, and included the decay of the dark boson to light hadronic states to calculate its width. We implemented a fit to the binned 
  $b \to s \ell^+ \ell^-$ data to find the preferred values of the mass and mixing parameters.

 We find that the base model with no direct $Z_D$ couplings to charged leptons is ruled out by constraints from low energy experiments.
 Two extensions of the model with direct muon and electron couplings provide an improvement in the fit to $b \to s$ data over both the SM and base model without direct couplings.  However, the many constraints permit only one of the extensions, with the viable parameter space confined to a small region. Consistency with measurements of the anomalous magnetic moment of the muon requires yet another source of new physics. We summarize the main results for each case below.
\begin{itemize}
    \item {\bf Case A: } For the base $Z_D$ model, we find that $\eps_Z \sim 10^{-3}$ and $M_{Z_D} \lsim 30$~MeV is required to explain the binned $b \to s \ell \ell$ branching fraction data; see Fig.~\ref{fig:CaseA}. However, the parameter space is excluded primarily by measurements of the proton and cesium weak charges in atomic parity violation experiments. Above $M_{Z_D} = 30$~MeV, the mixing parameters are severely constrained by FCNC measurements to which $Z_D$ contributes as a sharp resonance.
    \item {\bf Case B: } Case A is extended with a direct coupling of $Z_D$ with muons. To respect bounds on $\eps$ and $\eps_Z$ from meson mixing and APV, we fix $\eps = 10^{-4}$ and $\eps_Z = 10^{-5}$. As a result, the parameter space is restricted to $M_{Z_D} < 30$~MeV. The coupling $g_D^\mu \gsim 0.1$ provides an unsuppressed contribution to the branching fraction to the muon modes. The entire parameter space is ruled out because of enhancements to $K \to \mu \nu X$ and to the $W$ boson width.
    \item {\bf Case C: }
    In addition to a direct muon coupling, $Z_D$ has a fine-tuned direct coupling to electrons to cancel its coupling to electrons through mixing. This avoids the constraints from APV and $K \to \mu \nu e^+ e^-$. A fit to the $b \to s \mu^+ \mu^-$ observables gives a best fit at $M_{Z_D} = 30.2$~MeV and $g_D^\mu = 0.033$ for $\eps = \eps_Z = 10^{-4}$ with a pull of 4.94 from the SM. Bounds from neutrino trident production at CCFR, LHCb dark photon searches, $W$ width measurements and $K \to \mu \nu X$ rule out much of the allowed parameter space as shown in Fig.~\ref{fig:CaseC}. A $2\sigma$ region around $100 \lsim M_{Z_D} \lsim 200$ MeV and $0.015\lsim g_D^\mu \lsim 0.03$ remains viable provided a fine-tuned cancellation with other new physics is arranged to satisfy the constraint from the $a_\mu$ measurement.

\end{itemize}

\bigskip
\noindent
{\bf Acknowledgments}: 
We thank W.~Altmannshofer and J.~Dror for useful discussions, and J.~Kumar for technical help on \texttt{flavio}. A.R. acknowledges the hospitality of the Department of Physics, University of Basel where his visit was supported through  the  SU-FPDC  Grant Program.
A.D. and L.M. are supported in part by the U.S. National Science Foundation under Grant No.~PHY-1915142. D.M. is supported in part by the U.S. Department of Energy under Grant No.~de-sc0010504.

\appendix
\section{Form factors}
\label{sec:FF}

The form factors (FFs) for transitions to the vector hadronic states, $B\rightarrow K^*,\; B\rightarrow \rho,\; B\rightarrow \omega,\; B_s \rightarrow \phi,\; B_s \rightarrow K^* $ are as below. The vector and tensor hadronic currents are  parameterized by seven FFs~\cite{Bharucha:2015bzk}:
 \begin{eqnarray}
 c_V   \langle K^*(p,\eta) | \bar s \gamma^\mu(1 \mp \gamma_5) b | \bar B(p_B)   \rangle
 &=&   
   P_1^\mu \; {\cal V}_1(q^2) 
   \pm P_2^\mu \,
  {\cal V}_2(q^2) \pm P_3^\mu \,  {\cal V}_3(q^2)  \pm P_P^\mu {\cal V}_P(q^2) 
   \,,\nonumber  \\[0.1cm]
  c_V  \langle K^*(p,\eta) | \bar s iq_\nu \sigma^{\mu\nu} (1 \pm \gamma_5) b | \bar B(p_B) \rangle
  &=& \ P_1^\mu  T_1(q^2)   \pm  P_2^\mu  T_2(q^2) \pm  P_3^\mu  T_3(q^2) 
   \,, 
   \label{eq:ffbasis}
\end{eqnarray}
where the Lorentz structures $P_i^\mu$ are defined by~\cite{Lyon:2013gba}
\begin{eqnarray}
\label{eq:Vprojectors}
P_1^\mu  &=&  2 \varepsilon^{\mu}_{\phantom{x} \alpha \beta \gamma} \eta^{*\alpha} p^{\beta}q^\gamma \,, \qquad   P_2^\mu = i \{(m_B^2- m_{K^*}^2) \eta^{*\mu} - 
(\eta^*\!\cdot\! q)(p+p_B)^\mu\} \,, \\
 P_3^\mu &=&  i(\eta^*\!\cdot\! q)\{q^\mu -  \frac{q^2 }{m_B^2 - m_{K^*}^2} (p+p_B)^\mu \}\,,  \qquad  P_P^\mu = i (\eta^* \cdot q) q^\mu   \,,
\nonumber 
\end{eqnarray}
where $\eta$ is the vector meson polarization and $\varepsilon_{0123}=+1$ defines the convention for the Levi-Civita tensor.
The factor $c_V$ accounts for the flavor content of 
particles: $c_V=\sqrt{2}$ for $\rho^0$,
$\omega$ and $c_V=1$ otherwise.\footnote{To be
  precise, $c_V$ is $\sqrt{2}$ for $\rho^0$ in
  $b\to u$ and for $\omega$, and  $-\sqrt{2}$ for $\rho^0$ in
  $b\to d$, with the flavor wave functions $\rho^0\sim (\bar u u -
  \bar d d)/\sqrt{2}$ and $\omega \sim (\bar u u +
  \bar d d)/\sqrt{2}$. We assume that $\phi$ is a pure $s\bar s$ state.}
 The parameterization in Eq.~\eqref{eq:ffbasis} makes the 
correspondence between vector and tensor FFs explicit.  
The correspondence between ${\cal V}_{P,1,2,3}$ and the more traditional FFs $A_{0,1,2,3}$ and $V$ 
is as follows:
 \begin{eqnarray}
& {\cal V}_P(q^2) =  -\frac{2 m_{K^*}}{q^2} A_0(q^2) \,,  \quad {\cal V}_1(q^2) =  -\frac{V(q^2)}{m_B+m_{K^*}} \,, \quad     {\cal V}_2(q^2) =    -\frac{A_1(q^2)}{m_B-m_{K^*}} 
\,, \nonumber  \\[0.1cm]
 & {\cal V}_3(q^2) =  \frac{m_B+m_{K^*}}{q^2}    A_1(q^2) -   \frac{m_B-m_{K^*}}{q^2}    A_2(q^2)  \equiv \frac{2 m_{K^*}}{q^2} A_3(q^2) \,.
 \label{eq:VAs}
\end{eqnarray}

The form factors $f_+^P$, $f_0^P$ and $f_T^P$ which are relevant for
the $B\to P$ transition, where $P$ denotes the pseudoscalar hadronic states $\pi$, $K$ or $\eta$,
are defined as 
follows~\cite{Ball:2004ye}:\footnote{The following  notation is frequently used in the 
literature: $f_+=F_1$ and $f_0=F_0$.} 
\begin{eqnarray}
\label{eq:fplus}
\langle P(p)|V^P_{\mu}|B(p_B) \rangle &=& \big\{(p+p_B)_{\mu}-
\frac{m_B^2-m_P^2}{q^2}\,q_{\mu}\big\}
f^P_+(q^2)+\big\{\frac{m_B^2-m_P^2}{q^2}\,q_{\mu}\big\}  
\,f^P_0(q^2)\,,\nonumber \\\label{eq:fT}
 \langle P(p)|J^{P,\sigma}_{\mu}|B(p_B)\rangle &=& 
\frac{i}{m_B+m_P}\big\{q^2(p+p_B)_{\mu}-
(m_B^2-m_P^2)q_{\mu}\big\} f_T^P(q^2,\mu)\,,
\end{eqnarray}
where $V^{\pi,\eta}_{\mu} = \bar{u} \gamma_{\mu} b$ is the
standard weak current, $V^K_\mu$ is given by 
$V^{K}_{\mu} = \bar s \gamma_{\mu} b$ and 
$J_{\mu}^{\pi(\eta),\sigma} = \bar{d} \sigma_{\mu\nu}q^{\nu} b$,
$J_{\mu}^{K,\sigma} = \bar s \sigma_{\mu\nu}q^{\nu} b$ are
penguin currents. The momentum transfer is given by $q=p_B-p$ and
the physical range in $q^2$ is \mbox{$0 
\leq q^2 \leq (m_B-m_{P})^2$.}

The hadronic current in the decay process $K^+ \rightarrow \pi^+ \nu \bar{\nu}$ is given by~\cite{Wu:2007yh}
\beq
\langle \pi^+(p_\pi) | \bar d \gamma^\mu(1 \mp \gamma_5) s | K^+(p_K)   \rangle = f_+ (q^2) (p_K + p_\pi)^\mu +  f_- (q^2) (p_K - p_\pi)^\mu\,,
\eeq
where $q^2 = (p_K - p_\pi)^2$ and $f_\pm (q^2)$ are the FFs~\cite{Gao:1999qn,Bijnens:2002mg},
\beq
f_\pm (q^2) = f_\pm (0) (1+\lambda_\pm (q^2 / m^2_{\pi^+}))\,,
\eeq
with $\lambda_-=0$, $f_- (0) = -0.332$, $f_+ (0) = 0.57$ \cite{Gao:1999qn} and
\beq
\lambda_+ (q^2 / m^2_{\pi^+}) = \frac{1}{m^2_{\pi}}\sqrt{(m_K^2+m_\pi^2-q^2)^2 - 4m_K^2 m_\pi^2}\,.
\eeq

The three different components to the decay processes $K_L^0 \rightarrow \pi^0 e^+ e^-$ and $K_L^0 \rightarrow \pi^0 \nu \bar{\nu}$ are: 1) a CP-conserving process which proceeds through two-photon exchanges, 2) an indirect CP-violating  effect proportional to the parameter $\eps$, and 3) the direct CP-violating effect which is manifested in the penguin reactions.
We consider the the contribution of the direct CP-violating effect to constrain the mixing parameters $(\varepsilon,\; \varepsilon_Z)$. Note that the amplitude is proportional to the imaginary components of the CKM matrix elements in the hadronic loop coefficients in Appendix~\ref{sec:loop-factors}.
The form factors that describe the direct CP-violating contribution, which is manifested in the penguin reactions, to the hadronic current $K_L^0 \rightarrow \pi^0$ can be found in \cite{Mescia:2006jd}
\begin{gather}
\langle\pi^{0}\left(  K\right)  |\bar{s}\gamma^{\mu}d|K^{0}_L\left(  P\right)
\rangle=\left(  \left(  P+K\right)  ^{\mu}f_{+}^{K^{0}%
\pi^{0}}\left(  z\right)  +\left(  P-K\right)  ^{\mu}f_{-}^{K^{0}\pi^{0}%
}\left(  z\right)  \right)  \;,\label{Eq4}\\
f_{-}^{K^{0}\pi^{0}}\left(  z\right)  =\frac{1-r_{\pi}^{2}}{z}\left(
f_{0}^{K^{0}\pi^{0}}\left(  z\right)  -f_{+}^{K^{0}\pi^{0}}\left(  z\right)
\right)  \,,\nonumber
\end{gather}
where $r_\pi=M_\pi / M_K$, $z=q^2 / M_{K^0}^2$. The slopes extracted from $K_{\ell3}$ decays by neglecting
isospin breaking are~\cite{ParticleDataGroup:2022pth,KTeV:2004ozu,NA48:2004jcz,Yushchenko:2003xz,KLOE:2006kms},
\begin{equation}
f_{0,+}\left(  z\right)  \equiv f_{0,+}^{K^{0}\pi^{0}}\left(  z\right)
=\frac{f_{+}\left(  0\right)  }{1-\lambda_{0,+}z}\,,~~\lambda_{0}=0.18\,,~~
\lambda_{+}=0.32\,, \label{Eq5}%
\end{equation}
in the pole parametrization. Accounting for isospin breaking in $\pi^{0}-\eta$
mixing, at zero momentum transfer one finds~\cite{Marciano:1996wy}, 
\begin{equation}
f_{+}\left(  0\right)  =\left(  1.0231\right)  ^{-1}f_{+}^{K^{0}\pi^{+}%
}\left(  0\right)  \approx0.939\;, \label{Eq6}%
\end{equation}
with the Leutwyler-Ross prediction $f_{+}^{K^{0}\pi^{+}}\left(  0\right)
=0.961\left(  8\right)  $ \cite{Leutwyler:1984je}, which is confirmed by lattice
studies~\cite{Becirevic:2004ya,Dawson:2005zv,JLQCD:2005rbf}. 

The hadronic current $K_S^0 \rightarrow \pi^0$ is given by~\cite{Buchalla:2003sj}
\begin{gather}
\langle\pi^{0}\left(  K\right)  |\bar{s}\gamma^{\mu}d|K^{0}_S\left(  P\right)\rangle= W_S(Z) \left(  P+K \right)^{\mu},
\end{gather}
where
\beq
 W_S(Z)= (a_S + b_S Z)\,,
\eeq
with $|a_S|=1.08^{+0.26}_{-0.21}$, $b_S/a_S=m_{K_S}^2/m_\rho^2$, and $z=q^2/m_{K_S}^2$.

\section{Hadronic loop contributions}
\label{sec:loop-factors}
The Wilson coefficients for the dark photon contributions in Eq.~(\ref{DarkPhotonZ}) for $b \to sZ_D$ are obtained using the {\tt Peng4BSM@LO} 
package~\cite{Bednyakov:2013tca}:
\begin{eqnarray}
\left(E^{0,A}_{c2,c3}\right)_{L}&=& 0\,,
\end{eqnarray}

\begin{eqnarray}
(E^{0,A}_{c2,c3})_{R}&=&0\,,
\end{eqnarray}

\begin{eqnarray}
(E^{2,A}_{c2,c3})_{L}&=& \sum_{j=2}^3 \frac{e^3  \varepsilon (V_{\text{CKM}}^{\text{j2}}){}^*  V_{\text{CKM}}^{\text{j3}} \delta_{c2\;c3}}{1152 \pi ^2 \left(x_1^2-1\right){}^4 \left(x_j^2-1\right){}^4 M_W^2 s_w^2}\nonumber\\
&&\left(4 \left(3 x_1^8-30 x_1^6+54 x_1^4-32 x_1^2+8\right) \left(x_j^2-1\right){}^4 \log \left(x_1\right)-\right.\nonumber\\
&&(x_1^2-1)\left(4 \left(x_1^2-1\right){}^3 \left(3 x_j^8-30 x_j^6+54 x_j^4-32 x_j^2+8\right) \log \left(x_j\right)+\right.\nonumber\\
&&(x_j^2-1)\left(25 x_j^4-19 x_j^6+x_1^6 \left(32 x_j^4-57 x_j^2+19\right)+\right.\nonumber\\
&&\left.\left.\left.\left(-32 x_j^6+75 x_j^2-25\right) x_1^4+\left(57 x_j^6-75 x_j^4\right) x_1^2\right)\right)\right)\,,
\end{eqnarray}

\begin{eqnarray}
(E^{2,A}_{c2,c3})_{R}&=&0\,,
\end{eqnarray}

\begin{eqnarray}
(M^{1,A}_{c2,c3})_{L} &=&\sum_{j=2}^3 -\frac{i e^3 m_s  \varepsilon (V_{\text{CKM}}^{\text{j2}}){}^*  V_{\text{CKM}}^{\text{j3}} \delta_{c2\;c3}}{384 \pi ^2 \left(x_1^2-1\right){}^4 \left(x_j^2-1\right){}^4 M_W^2 s_w^2}\nonumber\\
&& \left(-12 x_1^4 \left(3 x_1^2-2\right) \left(x_j^2-1\right){}^4 \log \left(x_1\right)+\right.\nonumber\\
&& (x_1^2-1)\left(12 \left(x_1^2-1\right){}^3 x_j^4 \left(3 x_j^2-2\right) \log \left(x_j\right)+\right.\nonumber\\
&& (x_j^2-1)\left(x_1^6 \left(-29 x_j^4+31 x_j^2-8\right)+x_j^2 \left(8 x_j^4+5 x_j^2-7\right)+\right. \nonumber\\
&& \left.\left.\left.\left(29 x_j^6-6 x_j^2-5\right) x_1^4+\left(-31 x_j^6+6 x_j^4+7\right) x_1^2\right)\right)\right)\,,
\end{eqnarray}

\begin{eqnarray}
(M^{1,A}_{c2,c3})_{R}&=&\sum_{j=2}^3 -\frac{i e^3 m_b  \varepsilon (V_{\text{CKM}}^{\text{j2}}){}^*  V_{\text{CKM}}^{\text{j3}} \delta_{c2\;c3}}{384 \pi ^2 \left(x_1^2-1\right){}^4 \left(x_j^2-1\right){}^4 M_W^2 s_w^2}\nonumber\\
&& \left(-12 x_1^4 \left(3 x_1^2-2\right) \left(x_j^2-1\right){}^4 \log \left(x_1\right)+\right.\nonumber\\
&& (x_1^2-1)\left(12 \left(x_1^2-1\right){}^3 x_j^4 \left(3 x_j^2-2\right) \log \left(x_j\right)+\right.\nonumber\\
&& (x_j^2-1)\left(x_1^6 \left(-29 x_j^4+31 x_j^2-8\right)+x_j^2 \left(8 x_j^4+5 x_j^2-7\right)+\right. \nonumber\\
&& \left.\left.\left.\left(29 x_j^6-6 x_j^2-5\right) x_1^4+\left(-31 x_j^6+6 x_j^4+7\right) x_1^2\right)\right)\right)\,.
\end{eqnarray}

The equations are simplified by using CKM matrix unitarity and introducing $x_j = m_j/M_W$ where $m_j$ are the masses of the internal quarks in the loop and $\delta_{{c2\;c3}}$ is the Kronecker delta function on the color states of the incoming and outgoing quarks. 

The Wilson coefficients for the dark $Z$ contributions in Eq.~(\ref{DarkPhotonZ}) for $b \to sZ_D$ are
\begin{eqnarray}
(E^{0,Z}_{c2,c3})_{L}&=&\sum_{j=2}^3  \frac{ e^3 \varepsilon_Z (V_{\text{CKM}}^{\text{j2}}){}^* V_{\text{CKM}}^{\text{j3}} \delta_{c2\;c3}}{64 \pi ^2 \left(x_1^2-1\right){}^2 \left(x_j^2-1\right){}^2 c_w s_w^3}\nonumber\\
&&\left(2 x_1^2 \left(3 x_1^2+2\right) \left(x_j^2-1\right){}^2 \log \left(x_1\right)+\right.\nonumber\\
&& (x_1^2-1)\left(-2 \left(x_1^2-1\right) x_j^2 \left(3 x_j^2+2\right) \log \left(x_j\right)+\right. \nonumber\\
&& \left.\left.\left(x_j^2-1\right) \left(\left(x_j^2-1\right) x_1^4-\left(x_j^4-6\right) x_1^2+x_j^2 \left(x_j^2-6\right)\right)\right)\right)\,,
\end{eqnarray}

\begin{eqnarray}
(E^{0,Z}_{c2,c3})_{R}&=&0\,,
\end{eqnarray}

\begin{eqnarray}
(E^{2,Z}_{c2,c3})_{L}&=&\sum_{j=2}^3  \frac{ e^3 \varepsilon_Z (V_{\text{CKM}}^{\text{j2}}){}^* V_{\text{CKM}}^{\text{j3}} \delta_{c2\;c3}}{2304 \pi ^2 \left(x_1^2-1\right){}^4 \left(x_j^2-1\right){}^4 c_w M_W^2 s_w^3}\nonumber\\
&& \left(4\log \left(x_1\right)\left(\left(3 x_1^8-36 x_1^6+36 x_1^4-64 x_1^2+16\right) s_w^2-3\left(4-12 x_1^2+\right.\right.\right.\nonumber\\
&& \left.\left.x_1^8 \left(1-s_w^2\right)-8 x_1^6 \left(1-s_w^2\right)+3 x_1^4 \left(8 \left(1-s_w^2\right)+3\right)\right)\right)\nonumber\\
&& \left(x_j^2-1\right){}^4-(x_1^2-1)\left(4 \left(x_1^2-1\right){}^3 \log \left(x_j\right)\right.\nonumber\\
&& \left(\left(3 x_j^8-36 x_j^6+36 x_j^4-64 x_j^2+16\right) s_w^2-3\left(4-12 x_j^2+\right.\right.\nonumber\\
&& \left.\left.x_j^8 \left(1-s_w^2\right)-8 x_j^6 \left(1-s_w^2\right)+3 x_j^4 \left(8 \left(1-s_w^2\right)+3\right)\right)\right)-\nonumber\\
&& (x_1^2-x_j^2) (x_j^2-1)\left(10-5 x_1^2-11 x_1^4-5 x_j^2-26 x_1^2 x_j^2+\right.\nonumber\\
&& 43 x_1^4 x_j^2-11 x_j^4+43 x_1^2 x_j^4-38 x_1^4 x_j^4+\nonumber\\
&& (1-s_w^2)\left(-58-73 x_j^2+29 x_j^4+x_1^2 \left(-145 x_j^4+422 x_j^2-73\right)+\right.\nonumber\\
&& \left.x_1^4 \left(14 x_j^4-145 x_j^2+29\right)\right)-s_w^2\left(58+23 x_j^2+9 x_j^4+\right.\nonumber\\
&& \left.\left.\left.\left.\left(50 x_j^4+31 x_j^2+9\right) x_1^4+\left(31 x_j^4-234 x_j^2+23\right) x_1^2\right)\right)\right)\right)\,,
\end{eqnarray}

\begin{eqnarray}
(E^{2,Z}_{c2,c3})_{R}&=&0\,,
\end{eqnarray}

\begin{eqnarray}
(M^{1,Z}_{c2,c3})_{L}&=&\sum_{j=2}^3  \frac{i e^3 m_s \varepsilon_Z (V_{\text{CKM}}^{\text{j2}}){}^* V_{\text{CKM}}^{\text{j3}} \delta_{c2\;c3}}{768 \pi ^2 \left(x_1^2-1\right){}^4 \left(x_j^2-1\right){}^4 c_w M_W^2 s_w^3}\nonumber\\
&& \left(-12 x_1^2 \log \left(x_1\right) \left(x_1^2 \left(-2 c_w^2+2 s_w^2-1\right)+x_1^4 \left(6 c_w^2-1\right)+1\right)\right.\nonumber\\
&& \left(x_j^2-1\right){}^4+(x_1^2-1)\left(12 \left(x_1^2-1\right){}^3 x_j^2 \log \left(x_j\right)\right.\nonumber\\
&& \left(x_j^2 \left(-2 c_w^2+2 s_w^2-1\right)+x_j^4 \left(6 c_w^2-1\right)+1\right)-\left(x_1^2-x_j^2\right)\nonumber\\
&& (x_j^2-1)\left(10-17 x_1^2+x_1^4-17 x_j^2+22 x_1^2 x_j^2+7 x_1^4 x_j^2+x_j^4+\right.\nonumber\\
&& 7 x_1^2 x_j^4-14 x_1^4 x_j^4+2s_w^2\left(2+5 x_j^2-x_j^4+x_1^4 \left(2 x_j^4+5 x_j^2-1\right)+\right.\nonumber\\
&& \left.x_1^2 \left(5 x_j^4-22 x_j^2+5\right)\right)+2c_w^2\left(-5+10 x_j^2+7 x_j^4+\right.\nonumber\\
&& \left.\left.\left.\left.\left(31 x_j^4-26 x_j^2+7\right) x_1^4-2 \left(13 x_j^4+4 x_j^2-5\right) x_1^2\right)\right)\right)\right)\,,
\end{eqnarray}

\begin{eqnarray}
(M^{1,Z}_{c2,c3})_{R}&=&\sum_{j=2}^3  \frac{i e^3 m_b \varepsilon_Z (V_{\text{CKM}}^{\text{j2}}){}^* V_{\text{CKM}}^{\text{j3}} \delta_{c2\;c3}}{768 \pi ^2 \left(x_1^2-1\right){}^4 \left(x_j^2-1\right){}^4 c_w M_W^2 s_w^3}\nonumber\\
&& \left(-12 x_1^2 \log \left(x_1\right) \left(x_1^2 \left(-2 c_w^2+2 s_w^2-1\right)+x_1^4 \left(6 c_w^2-1\right)+1\right)\right.\nonumber\\
&& \left(x_j^2-1\right){}^4+(x_1^2-1)\left(12 \left(x_1^2-1\right){}^3 x_j^2 \log \left(x_j\right)\right.\nonumber\\
&& \left(x_j^2 \left(-2 c_w^2+2 s_w^2-1\right)+x_j^4 \left(6 c_w^2-1\right)+1\right)-\left(x_1^2-x_j^2\right)\nonumber\\
&& (x_j^2-1)\left(10-17 x_1^2+x_1^4-17 x_j^2+22 x_1^2 x_j^2+7 x_1^4 x_j^2+x_j^4+\right.\nonumber\\
&& 7 x_1^2 x_j^4-14 x_1^4 x_j^4+2s_w^2\left(2+5 x_j^2-x_j^4+x_1^4 \left(2 x_j^4+5 x_j^2-1\right)+\right.\nonumber\\
&& \left.x_1^2 \left(5 x_j^4-22 x_j^2+5\right)\right)+2c_w^2\left(-5+10 x_j^2+7 x_j^4+\right.\nonumber\\
&& \left.\left.\left.\left.\left(31 x_j^4-26 x_j^2+7\right) x_1^4-2 \left(13 x_j^4+4 x_j^2-5\right) x_1^2\right)\right)\right)\right)\,.
\end{eqnarray}

The Wilson coefficients for $b \rightarrow d Z_D$ and $d \rightarrow s Z_D$ can be obtained by changing the CKM matrix elements and the masses of the incoming and outgoing quarks in the equations above.

\bibliographystyle{JHEP}
\bibliography{ref}

\end{document}